\documentclass[10pt,reqno]{amsart}
\input{header}
\begin{document}
\maketitle
\begin{abstract}
In this work, we propose a denoising diffusion generative model (DDGM) trained with healthy electrocardiogram (ECG) data that focuses on ECG morphology and inter-lead dependence. Our results show that this innovative generative model can successfully generate realistic ECG signals. Furthermore, we explore the application of recent breakthroughs in solving linear inverse Bayesian problems using DDGM. This approach enables the development of several important clinical tools. These include the calculation of corrected QT intervals (QTc), effective noise suppression of ECG signals, recovery of missing ECG leads, and identification of anomalous readings, enabling significant advances in cardiac health monitoring and diagnosis.
\end{abstract}
\section{Introduction}
\label{sec:introduction}
For nearly a century, the electrocardiogram (ECG) has been an indispensable diagnostic tool in cardiology, dating back to the pioneering invention of Waller and further refined by the groundbreaking advances of Einthoven. When analyzing an ECG, cardiologists focus primarily on two key aspects: the sequence of characteristic events — namely the P-wave, QRS complex and T-wave — and the specific morphology of each of these events. Analyzing these sequences provides important insights into the heart rhythm and sheds light on possible irregularities or arrhythmias. At the same time, the detailed examination of the individual wave morphologies provides information about the electrical activity and its propagation in the different heart chambers. This dual focus  drives advances in cardiac diagnostics and patient care.

Current research in machine learning in cardiology has primarily focused on the temporal features of electrocardiograms (ECG) for the diagnosis of cardiac arrhythmias; see \cite{hannun2019cardiologist}, \cite{khurshid2022ecg} and \cite{madan2022hybrid} and the references therein. However, a deeper understanding of electrical propagation in the heart, focusing in particular on the morphology of individual ECG events, offers significant opportunities. This facet of ECG analysis, which is critical to the interpretation of complex cardiac activity, has significant implications for the advancement of preventive medicine applications. By harnessing these morphological insights, we can potentially develop more sophisticated and predictive machine learning tools that improve early detection and prevention strategies in cardiovascular health.

In the field of preventive medicine, especially for conditions like sudden cardiac death (SCD), gathering data directly linked to the event is often a significant challenge. SCD is responsible for an estimated $10\%$ of adult deaths in both Europe and the United States, frequently triggered by rapid ventricular arrhythmias such as ventricular fibrillation or ventricular tachycardia (VF/VT), which are commonly linked to structural heart diseases \cite{haissaguerre2018localized}. Given the high stakes, accurate assessment is essential for the effective deployment of treatments, including implantable cardiac defibrillators. An efficient and pragmatic approach, circumventing the need for event-specific data, involves leveraging databases of healthy heart signals to identify anomalies, as explored in studies like \cite{wen_cinc2021,Kang_2022_cinc2021_iop}.

In recent years, remarkable progress has been made in the field of generative modeling. A variety of techniques have been devised to train generative models capable of producing highly realistic patterns from original datasets. This is particularly notable in the context of complex, high-dimensional data\cite{kingma2019introduction,kobyzev2020normalizing,gui2021review}. Among these, Denoising-Diffusion Generative Models (DDGMs) stand out for their effectiveness. DDGMs transform noise into data that closely resembles the original domain through a series of denoising steps. Notably, they achieve remarkable success in generating images and audio, offering a compelling alternative to the complexities associated with adversarial training; see \cite{sohl2015deep,song2020score,song2021denoising,song2021maximum,benton2022denoising}.

The use of generative models as informative priors for solving inverse Bayesian problems has recently attracted considerable interest, a trend that can be attributed to their remarkable flexibility; see \cite{arjomand2017deep,wei2022deep,su2022survey,kaltenbach2023semi,shin2023physics,zhihang2023domain,sahlstrom2023utilizing}. In this context, DDGMs have proven to be well suited  as priors in solving inverse problems \cite{song2021denoising,chung2023diffusion,song2022solving,kawar2022denoising,kawar2021snips}. In particular, recent contributions \cite{cardoso2023monte} and \cite{wu2023practical} have introduced novel methods to sample from the posterior distributions of inverse problems with DDGMs as priors that comes with a series of theoretical guarantees in realisitc scenarios.
\subsection{Contributions} 
\begin{itemize}
\item Our research introduces the first DDGM specifically trained to generate realistic, healthy multi-lead electrocardiogram (ECG) single beats, with a focus on capturing accurate ECG morphology. This represents a significant advance in ECG research and provides a powerful tool for studying and understanding the intricacies of heart electrical activity.
\item We have used this novel DDGM as a prior to address various clinical challenges, without further training. These include denoising ECG signals, replacing missing traces to reconstruct incomplete data, and detecting anomalies.
\end{itemize}

\subsection{Related Works}
\label{sec:rel_works}
Several generative models of ECG have been proposed in the literature. One of the main focuses in the literature is the use of various generative models to perform data augmentation for the training of deep learning models. Several types of generative models have been proposed, from simulation-based models \cite{que2022ecg}, Recursive Neural Networks \cite{rajan2018generative} and Generative Adversarial Networks \cite{golany2019pgans, golany2020simgans, golany2021ecg, shaker2020generalization, chen2019emotionalgan}. Although this line of research leads to significant improvements in the performance of classifiers on various challenging tasks, it does not exploit the full potential of generative models. Other generative models have been designed to solve specifically the denoising problem, such as \cite{singh2020new, Li2023DeScoD}. With the exception of \cite{shaker2020generalization}, the goal of most of the proposed generative models is to generate a time frame of the ECG that corresponds to multiple heartbeats -- instead of focusing on a window representing a single heartbeat. 

Recently, \cite{Edmondm2023synthetic, alcaraz2023diffusion} proposed DDGMs comparable to those proposed in this paper. \cite{Edmondm2023synthetic} focuses on the generation of a single healthy beat for a single ECG lead. \cite{alcaraz2023diffusion} generates a 10-second time frame and is conditioned to various ECG statements. Instead, we propose a model that focuses on a single beat for multiple ECG leads and show how this type of model can be used as a prior to solve various ECG reconstruction tasks relevant to different clinical tasks. However, it is important to note that both models proposed in \cite{Edmondm2023synthetic, alcaraz2023diffusion} can be adapted (albeit with a limitation in \cite{Edmondm2023synthetic}) as a prior for reconstruction problems using the methods that we present in the second part of the current work. 
\subsection{Notations}
\label{sec:notations}
We denote the set of real numbers as $\rset$ and the set of natural numbers as $\nset$. For any pair $(a, b) \in \nset^2$, the interval $\intvect{a}{b}$ is defined as $\{i \in \nset | a \leq i \leq b\}$. Random variables are represented by uppercase letters, such as $V$, and their specific values are denoted by lowercase letters, such as $v$. For a given $d \in \nset$, the multivariate Gaussian distribution with a mean vector $\mu \in \rset^{d}$ and a covariance matrix $\Sigma \in \rset^{d^2}$ is expressed as $\gauss(\mu, \Sigma)$. The probability density function (\pdf) of this Gaussian distribution at a point $v \in \rset^{d}$ is denoted as $\gauss(v; \mu, \Sigma)$. If $\mu$ is a probability law, then $\state \sim \mu$ means that $\state$ is a random variable that is distributed according to $\mu$.
\section{DDGM for ECG}
\label{sec:ddgm_ecg}
\subsection{ECG}
The standard 12-lead ECG setup includes nine electrodes divided into three limb electrodes and six precordial electrodes. The limbs electrodes are combined to generate the Wilson Central Terminal (WCT), which defines a reference potential. The difference between the WCT potential and the precordial electrodes generate the precordial leads V1--V6, while the difference between the limb electrodes and the WCT generates the augmented leads aVL, aVR, aVF. The other 3 leads of the standard 12-lead ECG (I, II and III) correspond to the difference of potential between each pair of limb electrodes.
\subsection{DDGM}
This section provides a concise overview of Denoising-Diffusion Generative Models (DDGM). We focus on  the variance exploding (VE) framework \cite{song2020score}. Detailed derivations are given in \cite{sohl2015deep}, \cite{ho2020denoising}, \cite{song2020score} and \cite{song2021denoising}.

Most generative models are designed to transform a reference distribution, denoted $\refmeas$, into a distribution that approximates the original data distribution, denoted $\datadistr$.
To define DDGM, we introduce:
\begin{figure}[]
    \centering
    \includegraphics[valign=T,width=.8\hsize]{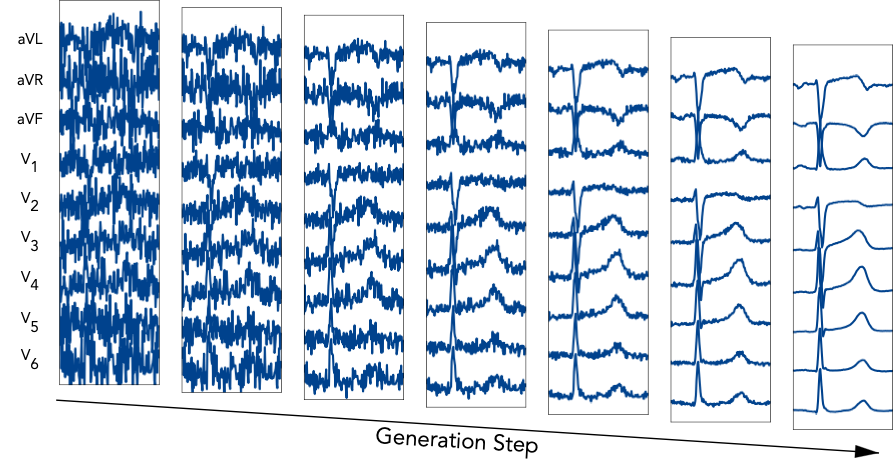}
    \caption{Example of single normal beat generated with DDGM, across multiple diffusion steps.}
    \label{fig:gen_ecg}
\end{figure}
\subsubsection{Forward noising process}
We define the reference distribution $\refmeas$ as $\gauss(0, \vefinalsigma^2 \idm)$, where $\vefinalsigma^2 >0$. The initial state of the forward process is sampled from $\state_0 \sim \datadistr$; independent noise with increasing variance is incrementally added to generate subsequent states
$\state_k = \state_{k-1} + \vestdker{k} \noise_{k}$,
where $k \in \nset$, $\vestdker{k} > 0$, and $\noise_{k} \sim \gauss(0, \idm)$.
The joint \pdf\ of the resulting Markov chain is
\begin{equation}
    \label{eq:forward_def}
    \fwmarg{0:\enddiff}{\chunk{\lstate}{0}{\enddiff}} = \datadistr(\lstate_0) \textstyle\prod_{k=1}^{\enddiff} \fwtrans{k}{\lstate_{k-1}}{\lstate_{k}}\eqsp,
\end{equation}
where the conditional \pdf\ of $\state_k$ given $\state_{k-1}$ is 
$\fwtrans{k|k-1}{\lstate_{k-1}}{\lstate_{k}} = \gauss(\lstate_{k}; \lstate_{k-1}, \vestdker{k}^2\idm)$.
Hence, the conditional \pdf\ of $\state_k$ given $\state_s$ for $k > s$ is
\begin{equation}
\label{eq:forward_def}
\fwmarg{k|s}{\lstate_k | \lstate_s} = \normdist(\lstate_s, (\stdexpl^2_k - \stdexpl^2_{s}) \idm) \eqsp,
\end{equation}
with $\stdexpl^2_k = \sum_{\ell=1}^{k}\varexpltrans_\ell$ (and $\stdexpl^2_0 = 0$). It is easy to see that if we chose $\enddiff \in \nset_{>0}$ such that $\stdexpl^2_{\enddiff} = \vefinalsigma^2$ and $\stdexpl^2_{\enddiff} \gg \lstate_0$, then $\fwmarg{\enddiff|0}{\cdot | \lstate_0}$ is close to $\refmeas$. Note that the forward kernel $\fwtrans{k}{}{}$ and forward marginals $\fwmarg{k|s}{}{}$ are coordinate wise-independent and we define 
\begin{align}
    \idxfwtrans{k}{\vecidx{\lstate}{k-1}{\ell}}{\vecidx{\lstate}{k}{\ell}}{\ell} &\eqdef \gauss(\vecidx{\lstate}{k}{\ell}; \vecidx{\lstate}{k-1}{\ell}, \vestdker{k}^2) \eqsp, \label{eq:forward_trans_cooord}\\ 
    \idxfwmarg{k|s}{\vecidx{\lstate}{s}{\ell}}{\vecidx{\lstate}{k}{\ell}}{\ell} &\eqdef \gauss(\vecidx{\lstate}{k}{\ell}; \vecidx{\lstate}{s}{\ell}, \stdexpl_{k}^2 - \stdexpl_{s}^2) \eqsp. 
    \label{eq:forward_marg_coord_2}
\end{align}
\subsubsection{Backward denoising process}
The backward denoising process reverts the forward process, starting from $\refmeas$.
 Adapting \cite{song2021denoising,song2021maximum} for the VE framework, we introduce an \emph{inference distribution} $\revbridge{1:\enddiff|0}{\lstate_0}{{\chunk{\lstate}{1}{\enddiff}}}$, depending on a sequence $\sequence{\infprocparam}[k][\nset]$ of hyperparameters satisfying 
$\infprocparam_{k}^2 \in [0,\stdexpl^2_k]$ for all $k \in \nset^*$, and defined as
\begin{align*}
\revbridge{1:\enddiff|0}{\lstate_0}{{\chunk{\lstate}{1}{\enddiff}}}&=
\revbridge{\enddiff|0}{\lstate_0}{\lstate_\enddiff}\prod_{k=\enddiff}^{2}
\revbridge{k-1|k, 0}{\lstate_k, \lstate_0}{\lstate_{k-1}} \eqsp,\\
\revbridge{\enddiff|0}{\lstate_0}{\lstate_\enddiff} = \gauss\left(\lstate_\enddiff ; \lstate_0, \vefinalsigma^2 \Id \right) \eqsp, &\quad
\revbridge{k-1|k, 0}{\lstate_k,\lstate_0}{\lstate_{k-1}} =\gauss\left(\lstate_{k-1} ; \boldsymbol{\mu}_{k-1}(\lstate_0,\lstate_k), \infprocparam_{k-1}^2 \idm_{d} \right) \eqsp,
\end{align*}
with
$\boldsymbol{\mu}_{k-1}(\lstate_0,\lstate_k)= \lstate_0 + (\stdexpl^2_{k-1} / \stdexpl^2_{k} - \infprocparam_{k-1}^2 / \stdexpl^2_{k})^{1/2} (\lstate_k - \lstate_0) \eqsp.$
For $k=\enddiff-1,\dots,1$, we define by backward induction the sequence 
\begin{equation*}
    \revbridge{k|0}{\lstate_0}{\lstate_k}
=  \int \revbridge{k|k+1,0}{\lstate_{k+1}, \lstate_0}{\lstate_{k}} \revbridge{k+1|0}{\lstate_0}{\lstate_{k+1}} \rmd \lstate_{k+1}\eqsp. 
\end{equation*}
The goal of the inference process is to provide a law that evolves backward conditionally on $\lstate_0$ and that matches the marginals $\{\fwmarg{k}{}\}_{k \in \intvect{1}{\enddiff}}$ of the forward process.
Indeed, as in \cite[Lemma 1]{song2021denoising}, it is possible to show that $\revbridge{k|0}{\lstate_0}{\lstate_k} =  \fwmarg{t|0}{\lstate_k|\lstate_0}$. The proof is postponed to the appendix. 

The backward process is derived from the inference distribution by replacing, for each $k \in [1:\enddiff-1]$, $\lstate_0$ in the definition $\revbridge{k|k+1, 0}{\lstate_{k+1}, \lstate_0}{\lstate_{k}}$ with a denoising network $\denoiser{0|k+1}[\theta](\lstate_{k+1}, \stdexpl_{k+1})$ trained to minimize
\begin{equation}
    \label{eq:opt_denoiser}
    \sum_{k=1}^{\enddiff} \gamma_{k}^2 \expec{\state_0 \sim \datadistr, \epsilon \sim \normdist(0, \idm)}{\|\denoiser{0|k}[\theta](\state_{0} + \stdexpl_k \epsilon, \stdexpl_k) - \state_{0} \|^2}\eqsp,
\end{equation}
where $\sequence{\gamma}[k][\intvect{1}{\enddiff}]$ is a sequence of weighting coefficients. 
Then, by replacing $\lstate_0$ by $\denoiser{0|k+1}[\theta](\lstate_{k+1})$ on $\revbridge{k|k+1, 0}{}{}$, we obtain the backward denoising kernels
\begin{equation*}
    \bwtrans{\theta}{k}{\lstate_{k+1}}{\lstate_k}  \eqdef \revbridge{k|k+1, 0}{\lstate_{k+1}, \denoiser{0|k+1}[\theta](\lstate_{k+1}, \stdexpl_{k+1})}{\lstate_k}
\end{equation*}
for $k \in \intvect{1}{\enddiff-1}$. By approximating $\revbridge{\enddiff|0}{\lstate_0}{\cdot}$ by $\refmeas$, we define the following joint law for the denoising process
\begin{equation*}
    \bwmarg{0:\enddiff}{\chunk{\lstate}{0}{\enddiff}} = \refmeas(\lstate_{\enddiff})\prod_{k=\enddiff-1}^{0}  \bwtrans{\theta}{k}{\lstate_{k+1}}{\lstate_k}
\end{equation*}
where for $k = 0$, we define $\bwtrans{\theta}{0}{\lstate_{1}}{\lstate_0}  \eqdef \normdist(\denoiser{0|1}[\theta](\lstate_{1}, \stdexpl_1), \infprocparam_0^2 \idm)$.

The evaluation phase of our generative model therefore consists of the following steps:
\begin{itemize}
    \item (i)  $\state_\enddiff \sim \refmeas$,
    \item (ii) $\state_{k} \sim \bwtrans{\theta}{k}{\lstate_{k+1}}{\cdot}$ for $k \in \intvect{1}{\enddiff-1}$,
    \item (iii) $\state_{0} \sim \normdist(\denoiser{0|1}[\theta](\lstate_{1}, \stdexpl_1), \infprocparam_0^2 \idm)$.
\end{itemize}
For a given choice of $\sequence{\gamma}[k][\intvect{1}{\enddiff}]$, it can be shown that \eqref{eq:opt_denoiser} is equivalent to minimizing the Kullback-Leibler divergence ($\kldivergence{}{}$)
between the inference process and the backward processes in the extended space of $(\state_0, \dots, \state_\enddiff)$. The precise definition of the $\kldivergence{}{}$ as well as the proof of the statement above is deferred to the appendix.
From now on, we drop the dependence on the parameter $\theta$ when referring to the generative model.
\subsection{The \model\ generative model}
Our goal is to develop a generative model to produce a single healthy ECG heartbeat, with  emphasis on the morphology of the QRS complex and T-wave. Essential to this objective is a standardized approach for extracting the heartbeat. From raw ECG strips, we first localized the R peaks, then ECGs were segmented into single-beat windows around the R peaks $(\operatorname{R-peak} - 192\operatorname{ms}, \operatorname{R-peak} + 512\operatorname{ms})$; see details in the supplement.  
Consequently, our focus shifts to the development of a DDGM designed to generate this specific time window.

We build a generative model using the augmented leads and leads V1 to V6. Given our primary concern with the morphology of each ECG lead and due to the fact that anomalies linked to the amplitude are easier to characterize, we opt to standardize the amplitude. We normalize each ECG lead window corresponding to the heartbeat by dividing it by the maximum absolute value attained during the QRS. The iterations of the generative model are illustrated in \cref{fig:gen_ecg}.
\subsection{Dataset and preprocessing}
\begin{figure}[H]
 \hspace{-2\tabcolsep}
\begin{minipage}[t]{0.18\textwidth}
\hspace{-2\tabcolsep}
\begin{tabular}{M{0.1\linewidth}@{\hspace{0\tabcolsep}} 
M{0.9\linewidth}@{\hspace{0\tabcolsep}}}
    \rotatebox[origin=c]{90}{\hspace{3.5\tabcolsep}Count}   &
\includegraphics[valign=C,width=\hsize]{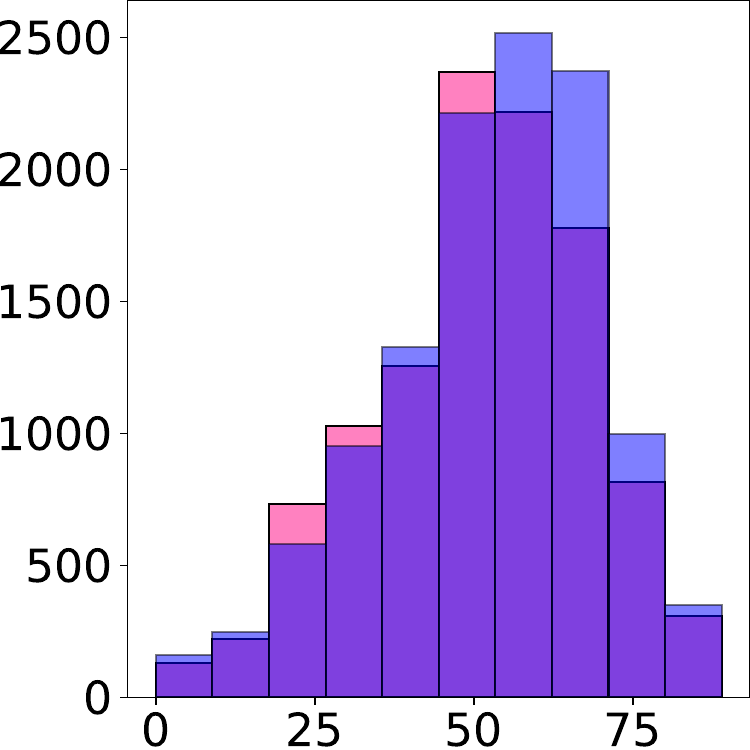}
  \\\addlinespace[-0.5ex]
  &
\hspace{2.7\tabcolsep}\small{Ages (train)}
\end{tabular}
\end{minipage}%
\quad
\begin{minipage}[t]{0.18\textwidth}
\hspace{-2.5\tabcolsep}
\begin{tabular}{M{0.9\linewidth}@{\hspace{0\tabcolsep}}}
\includegraphics[valign=C,width=\hsize]{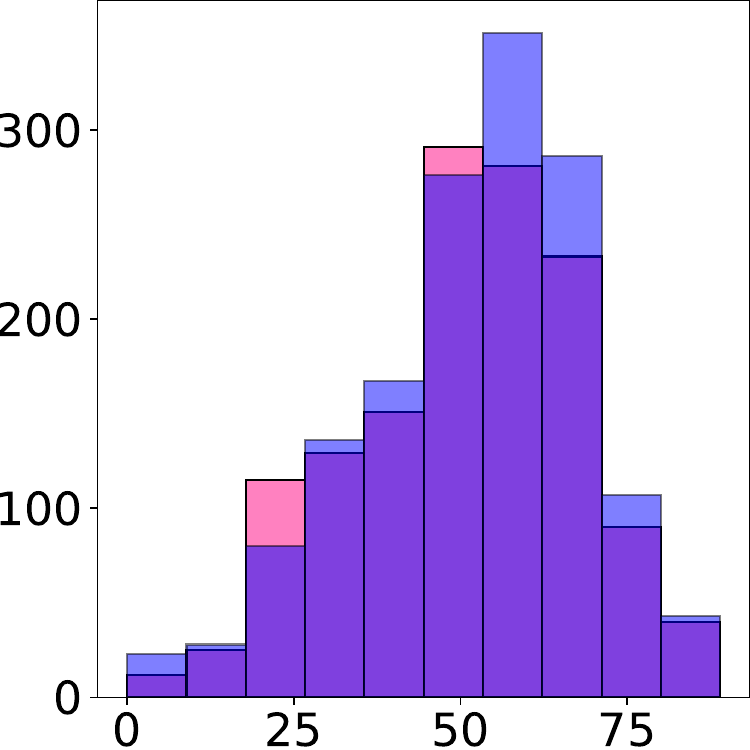}
  \\\addlinespace[-0.5ex]
\hspace{2\tabcolsep}\small{Ages (test)}
\end{tabular}
\end{minipage}%
\begin{minipage}[t]{0.18\textwidth}
\hspace{-3.5\tabcolsep}
\begin{tabular}{M{0.9\linewidth}@{\hspace{0\tabcolsep}}}
\includegraphics[valign=C,width=\hsize]{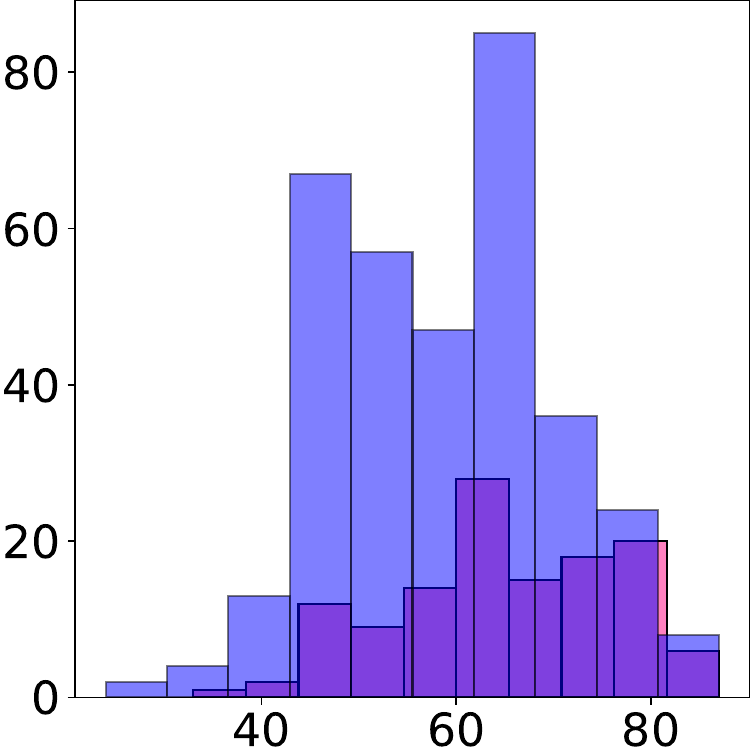}
  \\\addlinespace[-0.5ex]
\hspace{1.5\tabcolsep}\small{Ages (MI)}
\end{tabular}
\end{minipage}
\vspace{-1ex}
\caption{\small{Female (pink), male (blue) ages histograms in training (left), test (middle), MI (right) sets.}}\label{fig:dataset_age_by_sex}
\end{figure}

We use the 2021 PhysioNet Challenge \cite{PhysioNet,Reyna2021Willtwo,Reyna2022IssuesIT} which contains 43101 records from patients of different ages, genders and ethnic backgrounds (China, Russia, Germany, US) annotated with different ECG statements. 
The preprocessing of this data follows four stages. The sampling frequency of all the records is set to 250 Hz.
We then extract R peaks from the electrocardiogram data. In this phase, the first principal component is extracted channel-wise from the entire ECG. Subsequently, this extracted component is processed through a Savitzky-Golay filter, characterized by an order of $3$ and a window length of $15$. The extraction of R-peaks is then carried out based on the methodology proposed in \cite{Brammer2020}. Then, a window is extracted around each R-peak as described in the previous section. Therefore, we represent the state variable $\state$, corresponding to an ECG heart beat for the leads (aVL, aVR, aVF, V1--V6), as a $(9 \times 176)$ matrix. We could extract normal heart beats from a total of 28167 patients and Myocardial Infarctus (MI) heart beat from 468 patients.  See \cref{tab:dataset_splits} and \cref{fig:dataset_age_by_sex} for a detailed description of the division into training, cross-validation (CV) and test data, as well as the distribution of age and gender by group. Since recording time varies from patient to patient, and inter-patient variability is greater than inter-beat variability, a single beat is randomly sampled at when validating the model either in the training, CV, test or MI datasets.
\subsection{Training}
\label{sec:numerics:gen_model}
Various factors, such as age (A), sex (S), and the preceding R-R interval (RR), affect the morphology of the QRS complex and T-wave; see \cite{malik2013qt, salama2014sex, ball2014predicting}. We have thus developed a \emph{conditional} DDGM trained to generate the random variable $\state | A,S,RR$. From now on, except explicitly mentioned we will always generate an ECG $\state$ conditioned on $A,S,RR$.

The input of the $\denoiser{0|k}[\theta]$ is sampled from $\fwmarg{k}{}$ during training, as \eqref{eq:opt_denoiser}, thus its magnitude varies considerably with $k$. Therefore, using a neural network to model $\denoiser{0|k}[\theta]$ for all $k$ directly is a hard challenge. For this reason, following \cite{Karras2022edm}, we use the  reparametrization
    \begin{equation*}
        \label{eq:denoiser_reparam}
        \denoiser{0|k}[\theta](\lstate, \stdexpl_k) = \cskip(\stdexpl_k) \lstate +  \cout(\stdexpl_k) \fnet{\theta}(\cin(\stdexpl_k) \lstate, \cnoise(\stdexpl_k))\eqsp.
    \end{equation*}
where $\cin(\stdexpl_k) = (\stdexpl_k^2 + \stddata^2)^{-1/2}$, $\cskip(\stdexpl_k) = \cin(\stdexpl_k)^2\stddata^2$, $\cout(\stdexpl_k) = \stdexpl_k \stddata\cin(\stdexpl_k)$, $\cnoise(\stdexpl_k) = 4^{-1}\log(\stdexpl_k)$,
and $\stddata$ is a positive hyper parameter that corresponds to an estimate of the standard deviation of $\datadistr$.
Notice that for small $\stdexpl_k$, $\cskip(\stdexpl_k) \approx 1$ and $\cout(\stdexpl_k) \approx 0$, thus $\denoiser{0|k}[\theta](\lstate, \stdexpl_k) \approx \lstate$, which is expected since $\lstate$ is already a good reconstruction of the original data. On the contrary, if $\stdexpl_k$ is large, then $\cskip(\stdexpl_k) \approx 0$ and $\cout \approx 1$ thus $\denoiser{0|k}[\theta](\lstate, \stdexpl_k)$ relies heavily on the network $\fnet{\theta}$ to provide a good reconstruction.

The $\fnet{\theta}$ is inspired by the network proposed in \cite{dhariwal2021diffusion} that itself an enhancement of the original network firstly proposed in \cite{ho2020denoising}. Its main component are the U-Net blocks \cite{ronneberger2015u} which were originally designed for biomedical image segmentation.
The architecture of each U-Net block employs a sequence of residual layers and downsampling convolutions, succeeded by another sequence of residual layers featuring upsampling convolutions. Skip connections link layers of identical spatial dimensions. Additionally, a Multi-head attention layer \cite{vaswani2017attention} is applied to the output of each U-Net block with a number of heads corresponding to the original dimension divided by $64$.
We model the ECG data as a 1D time-series with $9$ channels, therefore we replace the original 2D convolutions by 1D convolutions in the U-Net blocks. All the convolutional layers kernels inside a U-Net block are of size $3$. The U-Net blocks can be stacked and a common usage in the literature is to combine several U-Net on different resolution levels, that are obtained by downsampling the data before feeding it to each block U-Net. We have experimented with using different resolution levels for the U-Net but found no significant gains w.r.t. using only one level (see appendix for a detailed discussion). 

Before feeding the input signal $\lstate$ through the U-Net block, we pass $\cin(\stdexpl_t)\lstate$ into a 1D convolutional layer with kernel-size 1 and $192$ channels, to obtain a $(192, 176)$-matrix $\operatorname{e}_\lstate$.
In order to incorporate information of the global position of $t \in \intvect{1}{176}$ that can be used by the 1D convolution layers and self-attention layers, we create a $(192, 176)$ positional encoding matrix where $\vecidx{\operatorname{e}}{\operatorname{time}}{\ell, t} \eqdef \vecidx{\posenc{t}}{}{\ell}$ where 
\begin{equation}
    \label{eq:positional_encoding}
    \vecidx{\posenc{t}}{}{\ell} =
    \begin{cases}
        \sin(1000^{-(r/96)} t) & \text{if} \quad \ell = 2r \eqsp, \\
        \cos(1000^{-(r/96)} t) & \text{if} \quad \ell = 2r + 1 \eqsp.
    \end{cases} 
\end{equation}
is the positional encoding proposed by \cite{vaswani2017attention}.
We also define $\operatorname{e}_{\stdexpl_k} \in \rset^{192}$ as the positional embedding of $\cnoise(\stdexpl_k)$, namely  $\vecidx{\operatorname{e}}{{\stdexpl_k}}{\ell} = \vecidx{\posenc{\cnoise(\stdexpl_k)}}{}{\ell}$ for $\ell \in \intvect{1}{192}$.
For the patient-specific conditional variables $A, S, RR$, we first start by normalizing $A$ and $RR$ as $\tilde{A} = (A - 50) / 50$, $\tilde{RR} = (RR - 400) / 400$. We use a one-hot encoding for $S$ to generate $\tilde{S} \in \{0, 1\}^2$. We concatenate $\tilde{S}, \tilde{A}, \tilde{RR}$ and apply a 2-layer dense net to obtain a $\operatorname{e}_{\operatorname{cat}} \in \rset^{192}$ embedding vector. Finally, we feed $\operatorname{e}_{\operatorname{time}} + \operatorname{e}_\lstate + \operatorname{e}_{\operatorname{cat}} + \operatorname{e}_{\stdexpl_k}$ to the U-Net block, where the $\rset^{192}$ vectors are transferred to a $(192 \times 176)$ matrix by repeating the vector over the second axis $176$ times. The input fed into a U-Net block therefore includes not only information about the original signal, complete with global time localization details, but also data about \( A \), \( S \), \( RR \) and the degree of distortion indicated by \( \stdexpl_k \). 

In each configuration examined, the model that achieved the lowest training loss in the cross-validation group (CV) was saved. There are several differences between the proposed model and \cite{alcaraz2023diffusion}. We are interested in modelling a fixed healthy heart beat, conditioned on patient information ($S, A$), while \cite{alcaraz2023diffusion} models a variable length ECG conditioned in several different ECG statements but that do not take into account patient information. Thus, the network \cite{alcaraz2023diffusion} has to account for the inter-beat variability and even handle different heart arrhythmias, thus justifying the utilisation of a different type of building block than the UNet proposed in our work. 
\subsection{Results}
 \begin{figure}[]
\begin{minipage}[t]{0.4\textwidth}
\hspace{-2.5\tabcolsep}
\begin{tabular}{M{0.15\linewidth}@{\hspace{-0.5\tabcolsep}} M{0.9\linewidth}@{\hspace{0\tabcolsep}}}
  \STAB{\rotatebox[origin=c]{90}{\small EMD\hspace{-2.5\tabcolsep}}}
  &  \includegraphics[width=\hsize]{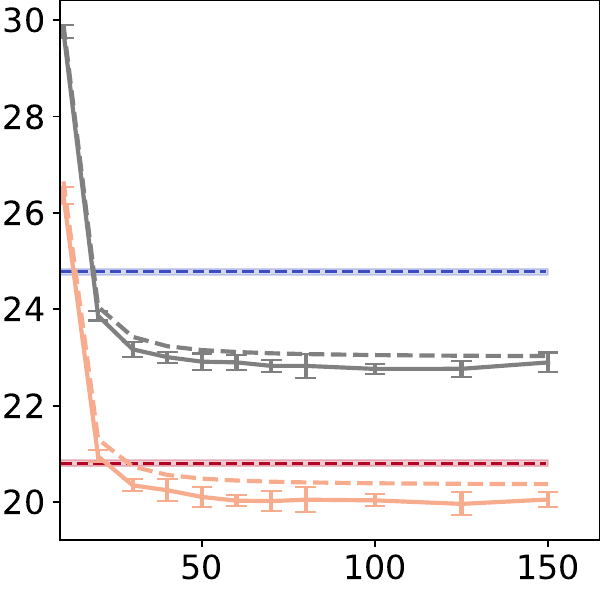}\\\addlinespace[-1ex]
  & \hspace{1\tabcolsep}\small Number of diffusion steps
\end{tabular}
\caption{
EMD of generated ECGs vs. test (dotted) and train (plain), w.r.t diffusion steps. Conditioned (resp. uncond.) DDGM in oragnge (resp. gray). EMD of test (resp. noisy-test) vs. train in red (resp. blue). Error bars correspond to different training batches of size  $2864$.
}\label{fig:emd_diffusion}
\end{minipage}%
    \quad
\begin{minipage}[t]{0.4\textwidth}
\hspace{-3\tabcolsep}
\begin{tabular}{M{0.15\linewidth}@{\hspace{0\tabcolsep}} 
M{0.9\linewidth}@{\hspace{0\tabcolsep}}}
    \rotatebox[origin=c]{90}{Uncertainty\hspace{-3\tabcolsep}}   &
\includegraphics[valign=C,width=\hsize]{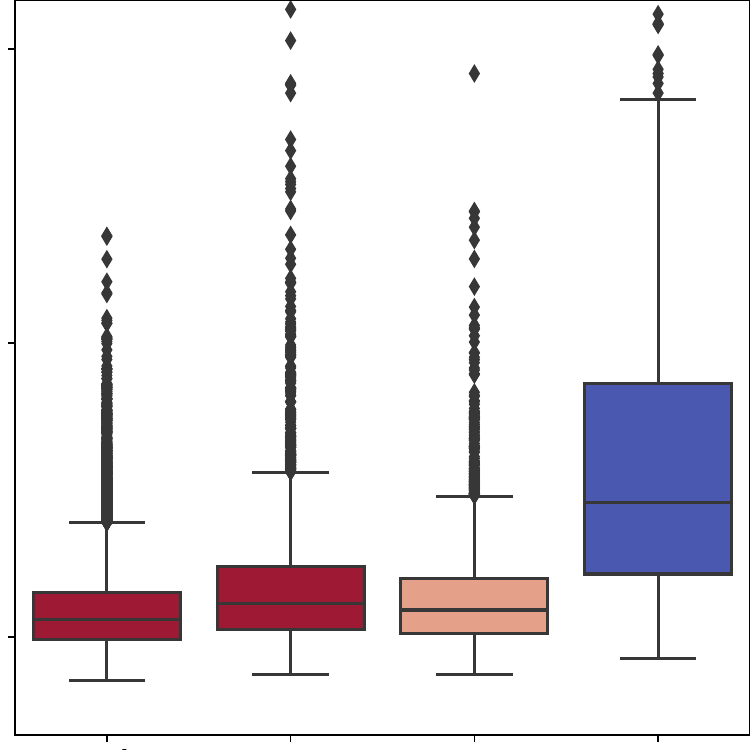}
  \\\addlinespace[-0.5ex]
  &
 \begin{tabular}{
M{0\linewidth}@{\hspace{-0.7\tabcolsep}}
M{0.25\linewidth}@{\hspace{-0.1\tabcolsep}}
M{0.25\linewidth}@{\hspace{0\tabcolsep}}
M{0.25\linewidth}@{\hspace{-0.1\tabcolsep}}
M{0.25\linewidth}@{\hspace{0.5\tabcolsep}}}
& \small{Train} & \small{Test} & \small{Gen} & \small{MI}
\end{tabular}
\end{tabular}
\caption{Box-plot of uncertainties estimation for train, test, generated (Gen) and MI heart beats.}\label{fig:uncertainty}
\end{minipage}
\end{figure}
In our study, we propose two novel metrics to assess the quality of synthetically generated  ECGs. These are: (1) a global distribution distance, and (2) a specific prediction error for each generated ECG. The global distribution distance is evaluated using the Earth Mover's Distance (EMD) \cite{genevay2016stochastic}. EMD effectively quantifies the dissimilarity between the predicted and target distributions by calculating the minimal transport cost required to transform the predicted distribution into the target distribution.
For several backward iterations $\enddiff$ varying between 2 and 150, we generate the same number of samples (2864) as the test set from a DDGM trained as described above, using the same $A, S, RR$ features as in the test set, as well as an unconditioned DDGM. The EMD is calculated from the generated set to the test set and the training set. To obtain comparable orders of magnitude, the training set was divided into batches of the same size (2864). The transport cost is defined as the $L^2$-distance over concatenated ECGs with $A, S, RR$ features to penalize the transport of an ECG to ECGs with different $A$, $S$, $RR$ features. To obtain comparable EMD for both DDGMs, unconditioned ECGs with $A, S, RR$ features randomly selected from the test set are also concatenated. \Cref{fig:emd_diffusion} shows that few diffusion steps are sufficient to generate convincing patterns and that conditioning over $A, S, RR$ leads to a smaller EMD. 
We also quantified the uncertainty of each predicted ECG, i.e. "how far" generated ECGs fall outside the support target distribution, following~\cite{Ciosek2020ConservativeUE}.
A randomly initialized network generates a 1024 representation vector, and a second network is trained to match this output by minimizing the distance between the vectors. The architecture of \cite{Ciosek2020ConservativeUE} is used, with a ResNet (known for its efficiency in ECG classification~\cite{ribeiro2020automatic}) for both networks, provided with linear layers for the trained network. To increase robustness, we use $B=10$ pairs of different networks, averaging the $L^2$ distances obtained for each pair over these bootstraps. The overall model is trained for 20 epochs over the 22580 training patients.
\Cref{fig:uncertainty} shows that the uncertainty of the 2864 ECGs generated with 50 diffusion steps closely matches that of the test set, in contrast to the MI ECGs.
\section{DDGM as a prior for ECG reconstruction}
\label{sec:DDGM:cond_gen}
We now consider the use of the trained DDGM as a prior for various reconstruction tasks.
We consider a partial observation $\lmeas \in \rset^{\dimys} \times \rset^{\dimyt}$ of $\dimys$ leads and $\dimyt$ temporal indices; defined by $\indmapl: \intvect{1}{\dimys} \rightarrow \intvect{1}{9}$ and $\indmapt: \intvect{1}{\dimyt} \rightarrow \intvect{1}{176}$ are two injective functions that specify the lead and time indices observed by $\lmeas$, respectively. 
Define by $\bwmarg{k}{}$ the marginal distribution of the backward process  $\bwmarg{k}{\lstate_k} \eqdef \int \refmeas(\lstate_\enddiff)\prod_{j=k}^{\enddiff-1} \bwtrans{}{j|j+1}{\lstate_{j+1}}{\lstate_{j}}\rmd \chunk{\lstate}{k+1}{\enddiff}$ for $k \in \intvect{0}{\enddiff-1}$.
Let us consider the following statistical model
\begin{equation}
    \label{eq:measurements}
    \vecidx{\ormeas}{}{\ell, t} = \vecidx{\state}{}{\indmapl(\ell), \indmapt(t)} + \noisestd_{\ell} \epsilon_{\ell, t}\eqsp,
\end{equation}
for $(\ell, t) \in \intvect{1}{\dimys} \times \intvect{1}{\dimyt}$, where $\epsilon_{\ell, t} \sim \normpdf(0, 1)$, and $\state \sim \bwmarg{0}{}$.
In the following we assume that we have observed a realisation $\lmeas$ of $\ormeas$ according to \eqref{eq:measurements}.

The posterior \pdf\ of $\state | \lmeas, \chunk{\sigma}{1}{\dimys}$ is $\filter{0}{\lmeas}{\lstate} \propto \pot{0}{\lmeas}{\lstate}\bwmarg{0}{\lstate}$ where
\begin{equation}
    \label{eq:likelihood}
    \pot{0}{\lmeas}{\lstate} =  \prod_{\ell=1}^{\dimys}\prod_{t=1}^{\dimyt} \gauss(\vecidx{x}{}{\indmapl(\ell), \indmapt(t)}; \vecidx{\lmeas}{}{\ell, t}, \sigma_{\ell}^2) \eqsp.
\end{equation}
Notice that sampling from $\filter{0}{\lmeas}{}$ is not trivial, since we do not have access to the \pdf\ $\bwmarg{0}{}$. Multiple studies, including  \cite{chung2023diffusion,kawar2022denoising,cardoso2023monte} have proposed methods for sampling from $\filter{0}{\lmeas}{}$.

First consider the homogeneous case: $\dimys=9$ and $\dimyt = 176$, $\noisestd_{\ell} = \noisestd_{\ell'} = \noisestd$ for all $(\ell, \ell') \in \intvect{1}{\dimys}^2$. Namely, $\lmeas$ is an observation of $\state \sim \bwmarg{0}{}$ corrupted with a Gaussian noise of standard deviation $\noisestd$.
Assume that there is a $k \in \intvect{1}{\enddiff}$ that satisfies $\stdexpl_k = \noisestd$.
Since $\fwtrans{k|0}{\lstate}{\lmeas} = \normdist(\lmeas; \lstate, \stdexpl_k^2 \idm) = \normdist(\lmeas; \lstate, \noisestd^2 \idm)$, \eqref{eq:likelihood} can be written as $\pot{0}{\lmeas}{\lstate} = \fwtrans{k|0}{\lstate}{\lmeas}$, leading to
$\filter{0}{\lmeas}{\lstate} \propto \fwtrans{k|0}{\lstate}{\lmeas} \bwmarg{0}{\lstate}$.
If $\fwmarg{0:\enddiff}{} = \bwmarg{0:\enddiff}{}$, then $\filter{0}{\lmeas}{\lstate} \propto \bwtrans{}{0|k}{\lmeas}{\lstate}$ and sampling from $\filter{0}{\lmeas}{}$ can be done by simply executing the backward process from $k$ to $0$ starting with $\lmeas$.

We now consider the general case where $\dimys \leq 9$, $\dimyt \leq 176$ and the $\{\noisestd_{\ell}\}_{\ell \in \intvect{1}{\dimys}}$ are not necessarily equal.
For each $\ell \in \intvect{1}{\dimys}$, suppose there is a forward diffusion-step $\stdmeet_{\ell}$ that satisfies $\stdexpl_{\stdmeet_{\ell}} \approx \noisestd_{\ell}$. Then, as for the homogeneous case, but this time for each coordinate, we can write 
\begin{equation*}
     \pot{0}{\lmeas}{\lstate} =  \prod_{\ell=1}^{\dimys}\prod_{t=1}^{\dimyt} \idxfwmarg{\stdmeet_\ell | 0}{\vecidx{\lstate}{}{\indmapl(\ell), \indmapt(t)}}{\vecidx{\lmeas}{}{\ell, t}}{{\indmapl(\ell), \indmapt(t)}}\eqsp.
\end{equation*} 
As for the homogeneous case, we can interpret the posterior as giving high probability to $\lstate$ that render, for each coordinate $(\ell, t)$, $\vecidx{\lmeas}{}{\ell, t}$ likely when propagating $\vecidx{\lstate}{}{\indmapl(\ell), \indmapt(t)}$ through the forward process $\idxfwmarg{\stdmeet_\ell | 0}{\vecidx{\lstate}{}{\indmapl(\ell), \indmapt(t)}}{\cdot}{{\indmapl(\ell), \indmapt(t)}}$.
Therefore, as for the homogeneous case, it can be shown (see \cite[Proposition 2.3]{cardoso2023monte}) that in the ideal case $\fwmarg{0:\enddiff}{} = \bwmarg{0:\enddiff}{}$ one can sample from $\filter{0}{\lmeas}{}$ by sampling from the backward process and enforcing $\vecidx{\state}{\stdmeet_\ell}{\indmapl(\ell), \indmapt(t)} = \vecidx{\lmeas}{}{\ell, t}$ for each coordinate $(\ell, t)$.

In \cite{cardoso2023monte}, the proposed method to sample from $\filter{0}{\lmeas}{}$ relies on the intuition presented above and use Sequential Monte Carlo (SMC) to sample from a sequence of distributions $\filter{k}{\lmeas}{\lstate} \propto \pot{k}{\lmeas}{\lstate} \bwmarg{k}{\lstate}$ which combine the backward process $\bwmarg{k}{}$ and a guiding function $\pot{k}{\lmeas}{}$ defined as:
\begin{equation}
    \label{eq:potential_smc}
    \pot{k}{\lmeas}{\lstate}=\prod_{\ell \in \biggervarset_k}\prod_{t=1}^{\dimyt} \gauss(\vecidx{x}{}{\indmapl(\ell), \indmapt(t)}; \vecidx{\lmeas}{}{\ell, t}, \stdexpl_k^2 - (1 - \delta)\sigma_{\ell}^2),
\end{equation}
with $\biggervarset_k \eqdef \{\ell \in \intvect{1}{\dimys} | k \geq \stdmeet_\ell\}$ and an hyper-parameter $\delta$. SMC consists in sampling recursively from a sequence of distributions $\{\filter{k}{\lmeas}{}\}_{k\in \intvect{0}{\enddiff}}$ that starts from an easy to sample initial distribution $\filter{\enddiff}{\lmeas}{}$ and progressively approaches $\filter{0}{\lmeas}{}$ through $\filter{k}{\lmeas}{}$.
Before delving into the details of the proposed SMC algorithm (\cref{subsec:smc}), it is important to contextualize the guiding functions $\pot{k}{\lmeas}{}$ and the intuition over $\pot{0}{\lmeas}{}$ given above.

For each coordinate $(\ell, t)$, the sequence $\{\gauss(\vecidx{x}{}{\indmapl(\ell), \indmapt(t)}; \vecidx{\lmeas}{}{\ell, t}, \stdexpl_k^2 - (1 - \delta)\sigma_{\ell}^2)\}_{k \geq \stdmeet_\ell}$ forms a sequence of Gaussian \pdf\ centered at $ \vecidx{\lmeas}{}{\ell, t}$ with decreasing standard deviation as $k$ goes to $\stdmeet_\ell$. This corresponds to guiding the backward process $\bwmarg{k}{}$ to enforce $\vecidx{\state}{\stdmeet_\ell}{\indmapl(\ell), \indmapt(t)} = \vecidx{\lmeas}{}{\ell, t}$.
Then, for $k < \min_{l\in\intvect{1}{\dimys}}\stdmeet_\ell$,  $\pot{k}{\lmeas}{}=1$ and the backward process alone is used to go from $\filter{k}{\lmeas}{}$ to the posterior $\filter{0}{\lmeas}{}$.

\begin{figure}[]
\centering
\includegraphics[valign=T,width=.775\hsize]{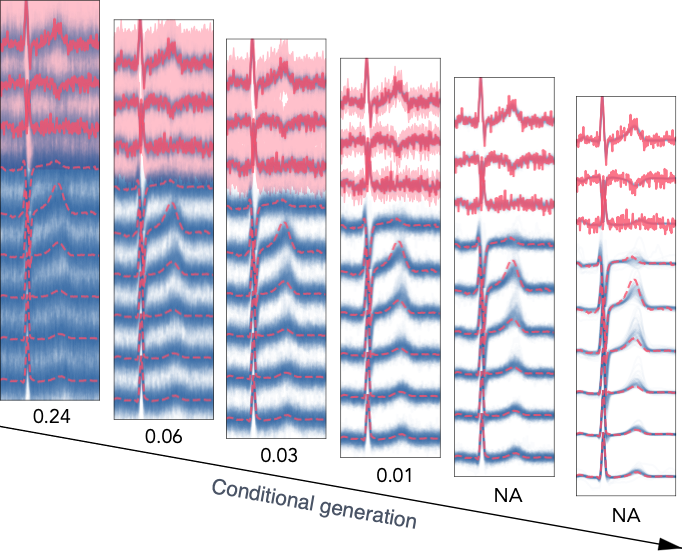}
\caption{Conditional generation example. Observation: (aVL, aVR, aVF) with $\noisestd=0.1$. Red solid/dashed lines: observed/real signal. Shaded zone: observed signal plus $3\times \text{std}$ of the guiding function \ref{eq:potential_smc}, std values on top. Blue: posterior samples.
}
\label{fig:cond_gen_ecg}
\end{figure}

The motivation behind the choice of the form of the variances in \eqref{eq:potential_smc} come from the forward process as well. Let $\vecidx{\trmeas}{\stdmeet_{\ell}}{\ell, t} = \vecidx{\lmeas}{}{\ell, t}$ and for $k > \stdmeet_\ell$, $ \vecidx{\trmeas}{k}{\ell, t} \sim \idxfwtrans{k}{\vecidx{\trmeas}{k-1}{\ell, t}}{\cdot}{\indmapl(\ell), \indmapt(t)}$. 
$\{\vecidx{\trmeas}{k}{\ell, t}\}_{k\geq \stdmeet_\ell}$ is a forward trajectory starting at value $\vecidx{\lmeas}{}{\ell, t}$ at iteration $\stdmeet_\ell$.
The role of the guiding function $\pot{k}{\lmeas}{}$ defined by \cref{eq:potential_smc} in $\filter{k}{\lmeas}{}$ is to make the trajectories $\vecidx{\trmeas}{k}{\ell, t}$ likely when sampling through the backward process $\bwmarg{k}{}$.
Indeed, for each coordinate $(\ell, t)$ and $k > \stdmeet_\ell$, assuming $\noisestd_{\ell} = \stdexpl_{\stdmeet_\ell}$, the density of $\vecidx{\trmeas}{k}{\ell, t}$ is (using \eqref{eq:forward_marg_coord}) $\idxfwmarg{k|\stdmeet_\ell}{\vecidx{\lmeas}{}{\ell, t}}{\vecidx{\lmeas}{k}{\ell, t}}{\indmapl(\ell), \indmapt(t)} = \gauss(\vecidx{\lmeas}{k}{\ell, t}; \vecidx{\lmeas}{}{\ell, t}, \stdexpl_k^2 - \sigma_{\ell}^2)$ where we recognize the terms in the product in \eqref{eq:potential_smc} with $\delta=0$. While the following definition holds for $k > \stdmeet_\ell$, for $\stdmeet_\ell$ we would have a Gaussian with $0$ variance, thus justifying the introduction of the parameter $\delta$ in \eqref{eq:potential_smc}.
\Cref{fig:cond_gen_ecg} shows an example of conditional generation from a noisy ($\noisestd = 0.1$) observation of aVL, aVR, aVF ($\dimys=3$, $\dimyt = 176$,  $\indmapt(t) = t$, $\indmapl(\ell)=\ell$). At the beginning of the generation (first column), the generated samples (in blue) are scattered and the standard deviation of the guiding function is high. As generation progresses (from left to right), the standard deviation of the guiding function decreases until it approaches 0. From then on (last two columns), $k < \min_{l\in[1,3]}\stdmeet_\ell$ and samples are generated by the backward process, i.e., $\pot{k}{\lmeas}{} = 1$.
\subsection{Sequential Monte Carlo (SMC)}
\label{subsec:smc}
SMC recursively builds an empirical approximation $\filter{k}{\npart, \lmeas}{} \eqdef \sum_{j=1}^{\npart} \omega_j \delta_{\particle_{k}^{j}}$ of the laws $\filter{k}{\lmeas}{}$ through a set of $\npart \in \nset_{>0}$ particles $\pchunk{\particle_k}{1}{\npart}$ and a discrete probability law $\{\omega_{j}\}_{j = 1}^{\npart}$. 
To do so, a recursion that establishes a relation between $\filter{k}{\lmeas}{}$ and $\filter{k+1}{\lmeas}{}$ is needed. For the current choice of $\{\filter{k}{\lmeas}{}\}_{k \in \intvect{0}{\enddiff}}$, the following relation holds:
\begin{equation}\begin{aligned}
    \label{eq:smc_recursion}
    \filter{k}{\lmeas}{\lstate} &\propto \int \frac{\pot{k}{\lmeas}{\lstate}\bwtrans{}{k|k+1}{\lstate_{k+1}}{\lstate} }{ \pot{k+1}{\lmeas}{\lstate_{k+1}}} \filter{k+1}{\lmeas}{\lstate_{k+1}}\rmd \lstate_{k+1} \eqsp.
    \end{aligned}
\end{equation}
Plugging into \eqref{eq:smc_recursion} a particular approximation of $\filter{k + 1}{\lmeas}{}$ given by $\filter{k + 1}{\lmeas, \npart}{} =  \sum_{j=1}^{\npart} \omega_j \delta_{\particle_{k+1}^{j}}$ we obtain
\begin{equation}
    \label{eq:smc_recursion_part_approx}
     \filter{k}{\lmeas}{\lstate} \propto \sum_{j=1}^{\npart} \omega_j \frac{\pot{k}{\lmeas}{\lstate}\bwtrans{}{k|k+1}{\particle_{k+1}^{j}}{\lstate} }{ \pot{k+1}{\lmeas}{\particle_{k+1}^j}} \eqsp.
\end{equation}
There are several ways of generating a particle approximation for $ \filter{t}{\lmeas}{}$ using \eqref{eq:smc_recursion_part_approx}. One way may be to consider sampling i.i.d. $\npart$ particles $\pchunk{\particle_k}{1}{\npart}$ from the Gaussian mixture distribution $\{\tilde{\omega}_k \bwtrans{}{t|t+1}{\particle_{k+1}^{k}}{\cdot}\}_{k=1}^{\npart}$ where $\tilde{\omega}_k \eqdef  (\omega_k \pot{t+1}{\lmeas}{\particle_{k+1}^k}^{-1}) / \sum_{j=1}^\npart \omega_j \pot{t+1}{\lmeas}{\particle_{k+1}^j}^{-1}$ and defining the new set of weights $\omega_k = \pot{t}{\lmeas}{\particle_k^k} / \sum_{j=1}^{\npart}\pot{t}{\lmeas}{\particle_k^j}$. One of the main issues with this way of sampling, is that the components of the mixture that generate the new particles do not include any information on $\lmeas$.

To address this issue, in \cite{cardoso2023monte} the authors propose considering the auxiliary kernel
$$\bwopt{k}{\lmeas}{\particle^{j}_{k+1}}{\lstate} \eqdef \pot{k}{\lmeas}{\lstate}\bwtrans{}{k|k+1}{\particle^{j}_{k+1}}{\lstate} / \operatorname{L}_k(\particle^{j}_{k+1}, \lmeas)$$ where $\operatorname{L}_k(\particle, \lmeas) \eqdef \int \pot{k}{\lmeas}{\lstate}\bwtrans{}{k|k+1}{\particle}{\lstate}  \rmd \lstate$. Therefore, by defining $\uweight{}{k}(\particle_{k+1}^j) \eqdef \operatorname{L}_k(\particle_{k+1}^j, \lmeas) / \pot{k+1}{\lmeas}{\particle_{k+1}^j}$ we can rewrite \eqref{eq:smc_recursion_part_approx} as
\begin{equation}
    \label{eq:smc_recursion_part_approx_mcg_diff}
    \filter{k}{\lmeas}{\lstate} \propto \sum_{j=1}^{\npart} \omega_j \uweight{}{k}(\particle_{k+1}^j) \bwopt{k}{\lmeas}{\particle^{j}_{k+1}}{\lstate} \eqsp.
\end{equation}
Thus, to obtain a particular approximation of $\filter{t}{\lmeas}{}$ we can sample $\npart$ ancestors
$$\{\anc{j}{t}\}_{j \in \intvect{1}{\npart}} \sim \categorical\big(\{ \omega_j \uweight{}{k}(\particle^{j} _{k+1}) / \sum_{i = 1}^{\npart} \omega_i \uweight{}{k}(\particle^{i} _{k+1})\} _{j = 1} ^{\npart} \big)^{\times \npart}$$
and then obtain the particles through the proposition kernels $\particle^{j}_{k} \sim \bwopt{k}{\lmeas}{\particle^{\anc{j}{k}}_{k+1}}{\cdot}$, leading to $\filter{k}{\npart, \lmeas}{} = \npart^{-1}\sum_{j=1}^{\npart} \delta_{\particle_{k}^{j}}$. Note that if we start $\filter{\enddiff}{\npart, \lmeas}{} = \npart^{-1} \sum_{j=1}^{\npart}  \delta_{\particle_{\enddiff}^{j}}$ then we have $\filter{k}{\npart, \lmeas}{} = \npart^{-1} \sum_{j=1}^{\npart}  \delta_{\particle_{k}^{j}}$ for all $k \in \intvect{0}{\enddiff}$.
Therefore, we drop the $\{\omega_j\}_{j=1}^{\npart}$ as they are constant.
The resulting algorithm is detailed in \cref{alg:guided_filter} and the explicit expression for both $\bwopt{k}{\lmeas}{}{}$ and $\uweight{k}{}$ are obtained using conjugate formula as follow.
\begin{algorithm2e}
    \caption{SMC}
    \label{alg:guided_filter}
    \SetKwInOut{KwInput}{input}\SetKwInOut{KwOutput}{output}
    \KwInput{Number of particles $\npart$, observation $y$}
    \KwOutput{$\pchunk{\particle _0}{1}{\npart}$}
    \tcp{\small{Operations involving index $i$ are repeated for each $i \in [1:\npart]$}}
    \BlankLine
    $\particle^i _\enddiff\sim \normpdf(\mathbf{0}_{d_x}, \vefinalsigma^2\idm_{d_x})$;\\
    \BlankLine
    \For{$k \gets \enddiff-1:0$}{
            \BlankLine
        $\anc{i}{k} \sim \categorical\big(\{ \uweight{}{k}(\particle^{j} _{k+1}) / \sum_{i = 1}^{\npart} \uweight{}{k}(\particle^{i} _{k+1})\} _{j = 1} ^{\npart} \big)$;\\
            \BlankLine
        $\particle^i_k \sim \bwopt{k}{\lmeas}{\particle^{\anc{i}{k}}_{k+1}}{\cdot}$;}
\end{algorithm2e}
\begin{align*}
\bwopt{k}{\lmeas}{\lstate_{k+1}}{\lstate_k} &= 
\prod_{\ell \notin \indmapl(\biggervarset_k)}
\prod_{t \notin \indmapt(\intvect{1}{\dimyt})}\normpdf(\vecidx{\lstate}{k}{\ell, t}; \vecidx{\mu_{k}(\lstate_{k+1})}{}{\ell, t}, \infprocparam_k^2) 
 \prod_{\ell \in \biggervarset_k} \prod_{k=1}^{\dimyt}
\normpdf(\vecidx{\lstate}{k}{\indmapl(\ell), \indmapt(t)}; \vecidx{\mu_{k, \lmeas}(\lstate_{k+1})}{}{\ell, t}, \frac{\infprocparam_k^2\sigma_{k,\lmeas}^2}{\infprocparam_k^2 + \sigma_{k, \lmeas}^2})\eqsp,\\
\uweight{\lmeas}{k}(\lstate_{k+1}) &= \prod_{\ell \in \biggervarset_k} \prod_{t=1}^{\dimyt}\smash{\gauss\left(\vecidx{\mu}{k}{\indmapl(\ell), \indmapt(t)}; \vecidx{\lmeas}{}{\ell, t}, \infprocparam_{k+1}^2 + \sigma_{k, \lmeas}^2\right)}
\big / \prod_{\ell \in \biggervarset_{k+1}}\prod_{t=1}^{\dimyt}\gauss(\vecidx{\lstate}{k+1}{\indmapl(\ell), \indmapt(t)}; \vecidx{\lmeas}{}{\ell, t}, \sigma_{k+1, \lmeas}^2) \eqsp,
\end{align*}
where $\sigma_{k, \lmeas}^2 \eqdef \stdexpl_k^2 - (1 - \delta)\noisestd_\ell^2$, $\mu_{k}\eqdef\boldsymbol{\mu}_{k}(\lstate_{k+1}, \denoiser{0|k+1}( \lstate_{k+1}))$ and
\begin{equation*}
    \vecidx{\mu_{k, \lmeas}(\lstate_{k+1})}{}{\ell, t} \eqdef \frac{\infprocparam_k^2\vecidx{\lmeas}{}{\ell, t} + \sigma_{k,\lmeas}^2\vecidx{\mu}{k}{\indmapl(\ell), \indmapt(t)}}{\infprocparam_k^2 + \sigma_{k,\lmeas}^2}\eqsp.
\end{equation*}
The resulting SMC algorithm comes with several statistical guarantees, in particular, a $\bigo(1 / \sqrt{\npart})$ bound on the mean squared error $\| \filter{0}{\npart, \lmeas}{h} - \filter{0}{\lmeas}{h} \|_2$ for every bounded function $h$. See \cite{cardoso2023monte} for further details.
\subsection{Parameter inference}
In the previous sections, we consider the parameters $\paramem \eqdef (\noisestd_{1}, \cdots, \noisestd_{\dimys})$ to be known. In several applications, this might not be true.
Therefore, in this section we propose a method to produce Maximum Likelihood Estimations (MLE) of $\paramem$.
Our goal is to find
\begin{equation}
\label{eq:mle_noise_std}
    \paramem^* \eqdef \argmax_{\paramem \in \rset^{\dimys} 
    }l(\paramem) \eqsp,
\end{equation}
where $l(\paramem) \eqdef \smash{\log \int \pot{0}{\lmeas, \paramem}{\lstate}\bwmarg{0}{\lstate}\rmd\lstate}$ and
we made explicit the dependency on the $\paramem$ parameter in $\pot{0}{\lmeas, \paramem}{}$. 
One way of tackling this problem is by gradient ascent. The gradient $\log \nabla_{\paramem} l(\paramem)$ can be written using Fisher's inequality as
\begin{align}
\label{eq:gradient_mle_noise}
    \nabla_{\paramem} l(\paramem) &= \int \nabla_{\paramem}\pot{0}{\lmeas, \paramem}{\lstate} \frac{\bwmarg{0}{\lstate}}{l(\paramem)}\rmd\lstate = \int \nabla_{\paramem} \log \pot{0}{\lmeas, \paramem}{\lstate} \frac{\pot{0}{\lmeas, \paramem}{\lstate}\bwmarg{0}{\lstate}}{l(\paramem)}\rmd\lstate = \int  \nabla_{\paramem} \log \pot{0}{\lmeas, \paramem}{\lstate} \filter{0}{\lmeas, \paramem}{\lstate} \rmd \lstate  \eqsp.
\end{align}
This integral is intractable, but the methodology developed in the previous section allows us to compute a Monte Carlo estimate of $\nabla_{\paramem} l(\paramem)$, by replacing $\filter{0}{\lmeas, \paramem}{}$ in \eqref{eq:gradient_mle_noise} by its particular approximation  $\filter{0}{\lmeas, \npart, \paramem}{} = \npart^{-1} \sum_{m=1}^{\npart} \delta_{\particle_{0}^m}$ given by \cref{alg:guided_filter} leading to
$\widehat{\nabla_{\paramem} l(\paramem)}[\pchunk{\particle_0}{1}{\npart}] \eqdef \npart^{-1} \sum_{m=1}^{\npart} \nabla_{\paramem} \log \pot{0}{\lmeas, \paramem}{\particle_{0}^m}$ which can be calculated analytically from \eqref{eq:likelihood}. We can also make this estimator more robust by running several parallel \cref{alg:guided_filter} and averaging the resulting estimators.
The algorithm used to solve \eqref{eq:mle_noise_std} is given in \cref{alg:mle_noise_std} and \cite[Section 11]{infhidden}.
\begin{algorithm2e}
    \caption{MLE for $\paramem$.}
    \label{alg:mle_noise_std}
    \SetKwInOut{KwInput}{input}\SetKwInOut{KwOutput}{output}
    \KwInput{Number of particles $\npart$, Number of parallel chains $N_c$, observation $y$, initial value $\paramem_0$, number of steps $\nmle$ and learning rate $\lrmle$.}
    \KwOutput{$\paramem_{\nmle}$}
    \For{$i \gets 0:\nmle-1$}{
            \BlankLine
            Sample $\pchunk{\particle_{0, i}}{1}{\npart N_c}$ by running $N_c$ parallel \cref{alg:guided_filter} with $\paramem_i, \npart$;\\
            Update $\paramem_{i+1} \eqdef \paramem_i + \frac{\lrmle}{ (i+1)^{0.6}} \widehat{\nabla_{\paramem} l(\paramem_i)}[\pchunk{\particle_{0, i}}{1}{\npart N_c}] $;}
\end{algorithm2e}
\subsection{Numerical evaluation}
We now focus on evaluating numerically the performance of using the trained DDGM as a prior for several reconstruction problems. It is important to focus on the fact that all the examples here do not involve any further optimization, they are achieved by sampling only.
In all the applications in this section, to sample from the posterior we start by first running \cref{alg:mle_noise_std} for a maximum of $\nmle$ to obtain an estimate that is then used to sample from the posterior $\state|\lmeas, \paramem_{\nmle}$ using \cref{alg:guided_filter}. 

\subsubsection{Denoising}\label{sec:numerics:applications:denoising}
We start by the most direct application, which the denoising with Gaussian noise. In this case, all the coordinates are observed ($\indmapl = \operatorname{id}, \indmapt = \operatorname{id}$) and the goal of the sampler is finding the distribution of possible signals given a noisy version $\lmeas$ of the same signal. To evaluate the performance of our algorithm, we sample for each patient an homogeneous time-wise noise but with a per-lead standard deviation $\noisestd_\ell \sim \exp(0.2)$.

From the posterior samples, we calculate the Mahalanobis distance, defined as 
\begin{equation}
    \label{eq:mahanalobis_def}
    \mhdist{\lstate}{\mu, \Sigma} \eqdef \sqrt{(\lstate - \mu)^T \Sigma^{-1} (\lstate - \mu)}\eqsp,
\end{equation}
where $\mu = \expec{}{\state}$ and $\Sigma = \expec{}{(\state - \mu)(\state - \mu)^T}$, and all the expectations are taking w.r.t the law of $\state|\lmeas, \paramem_{\nmle}$. 
In order to obtain an interpretable value for $\mhdist{}{}$, we divide it by the percentile $99\%$ of the $\chi^2$ distributions with $176$ degrees of freedom. We call it the rescaled Mahanalobis distance. This can be used to device an outlier detection tool, an generally it has been considered that values above $1$ are generally outlier \cite{manhalobis_normalisation}. The rescaled Mahanalobis distance, averaged over the ECG leads, is of $0.130 +/- 0.003$ over the whole test set. In order to asses as well the parameter estimation, we have calculated the total absolute deviation between the real $\paramem$ and $\paramem_{\nmle}$ obtained by \cref{alg:mle_noise_std}. We have obtained a value of $0.03 +/- 0.001$.
\begin{figure}[]
\centering
\includegraphics[valign=T,width=.6\hsize]{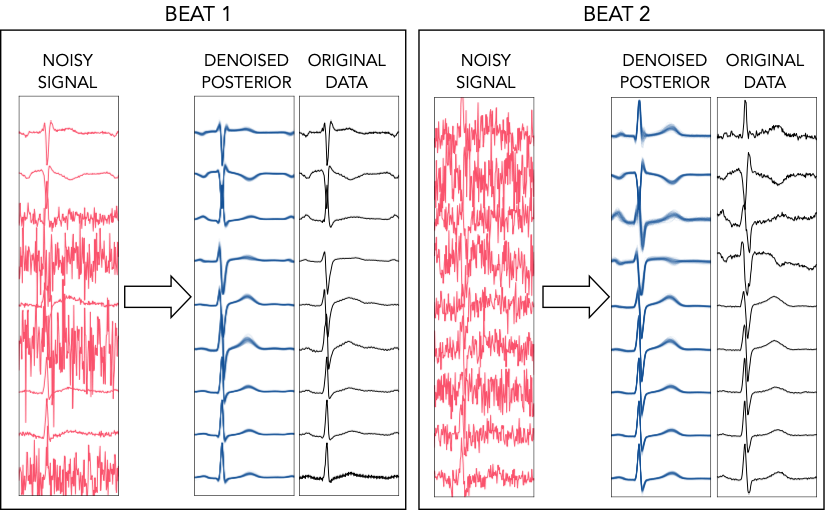}
\caption{Illustration of denoising with Gaussian noise for two different test signals. The original data used to generate the noised observation used to generate the denoised posterior samples is shown in black for illustration purposes.}
\label{fig:denoising}
\end{figure}
In \Cref{fig:denoising} we present two examples of signal denoising. These examples show that our approach effectively concentrates the posterior and approximates the original signal very well even in scenarios where the noise variance is large. However, it is important to note that a concentrated posterior around the original signal is not generally expected. This is especially true for certain inverse problems, which are inherently ill-posed and may have multiple solutions where the original signal is only one possible mode. Nonetheless, in the context of this example, it is valuable to illustrate the proximity of the posterior to the original signal.

\begin{minipage}[T]{0.45\textwidth}
\subsubsection{Missing leads}
We investigate the application of \algo\ to reconstruct missing leads in ECG data. In particular, we simulate scenarios with missing leads in the test series by withholding one lead from each test sample and trying to reconstruct it with \algo. We set $\nmle=10$. In this setup, $\ormeas$ represents all leads except the one to be reconstructed. For comparison, we also reconstruct the missing leads with Dower matrices \cite[Chapter 11, Vol 1]{macfarlane2010comprehensive}, where the comparative results are listed in \cref{tab:missing_leads:r2}.

We can see an almost perfect reconstruction of the leads using our method. Note that there is no point in trying to reconstruct aVL, aVR or aVF since not observing such a lead means that one of the limb electrodes is not available, which means that it is not possible calculate the WCT terminal and thus any of the precordial or augmented leads (aVL, aVR, aVF, V1, V2, V3, V4, V5, V6).
\begin{filecontents*}{data/missing_leads_r2.csv}
,track,mcg_diff,mcg_diff_num,dowers,dowers_num
3,V1,0.98 ± 0.00,0.9797339,0.71 ± 0.05,0.7095259
4,V2,0.99 ± 0.00,0.98965085,0.79 ± 0.05,0.7866694
5,V3,0.99 ± 0.00,0.9932454,0.76 ± 0.06,0.7560162
6,V4,0.99 ± 0.00,0.99378,0.87 ± 0.03,0.8686142
7,V5,0.98 ± 0.02,0.9794967,0.86 ± 0.08,0.8647288
8,V6,0.99 ± 0.01,0.9863399,0.86 ± 0.03,0.86001307
\end{filecontents*}
\DTLloaddb{missing_leads_r2}{data/missing_leads_r2.csv}
\begin{table}[H]
    \centering
    \resizebox{.6\textwidth}{!}{
    \begin{tabular}{|c|c|c|c|c|c|c|}%
        \hline
        Lead & \algo & Dowers \\
        \hline
        \DTLforeach*{missing_leads_r2}{
            \track=track,
            \distours=mcg_diff,
            \distoursnum=mcg_diff_num,
            \distdowers=dowers,
            \distdowersnum=dowers_num}{
            \track &
            \DTLgmaxall{\rowmax}{\distoursnum,\distdowersnum}%
            \dtlifnumeq{\distoursnum}{\rowmax}{\bf}{}\distours &
            \dtlifnumeq{\distdowersnum}{\rowmax}{\bf}{}\distdowers\DTLiflastrow{}{\\
            }}
        \\\hline
    \end{tabular}%
    }
    \vspace{-3mm}
    \caption{R2 score for each missing lead, using all the other leads. The confidence intervals correspond to CLT $95\%$ intervals over the whole test dataset.}
    \label{tab:missing_leads:r2}
\end{table}
\end{minipage}
\hspace{0.05\textwidth}
\begin{minipage}[T]{0.5\textwidth}
    \begin{figure}[H]\includegraphics[valign=T,width=1\hsize]{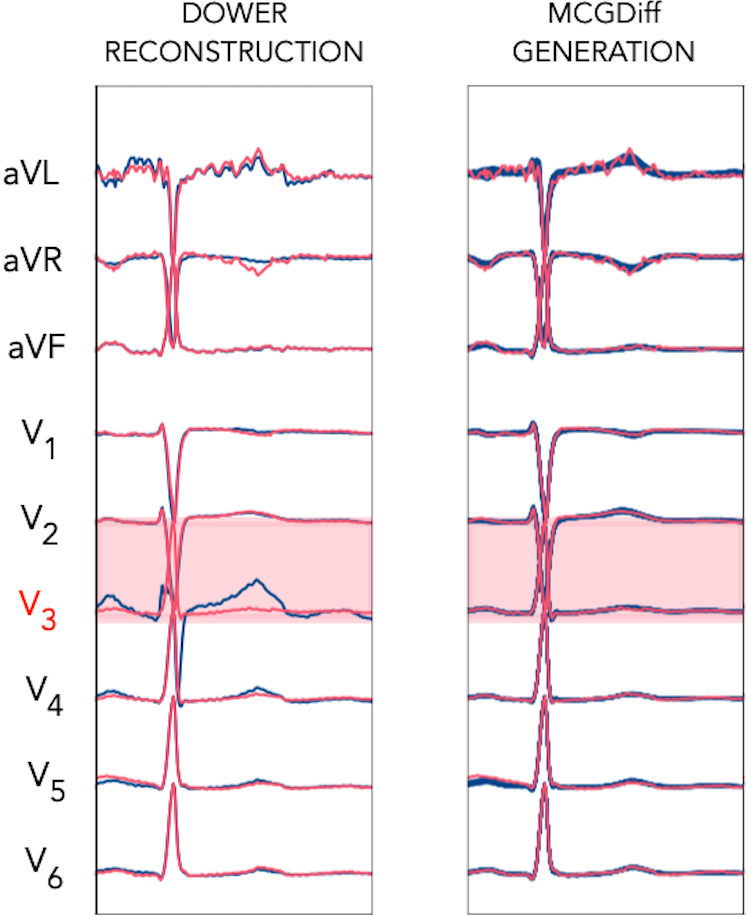}
    \vspace{-3mm}
    \caption{Illustration of reconstruction of V3 using the proposed method on the left and dowers matrix on the right.}
    \end{figure}
\end{minipage}

\subsubsection{QT vs RR}
The $\QT$ interval, a critical parameter in electrocardiography, is known to vary with the preceding $\RR$ interval. This relationship, which is crucial for the assessment of cardiac function, has been modeled by various empirical formulas. The best known of these are the formulas of Bazett \cite{bazett1997analysis}, Fridericia \cite{fridericia1921systolendauer} and Framingham \cite{sagie1992improved}. While these formulas provide general guidelines, it's important to note that their coefficients are patient-specific and indicate the variability of cardiac responses in different individuals. In this study, we aim to determine whether our model can accurately reproduce this $\QT$-$\RR$ relationship by conditioning on the observed QRS complex. Our approach is to generate conditional states $\state |\ormeas, \paramem_{\nmle}, \RR$, where $\ormeas$ spans the initial portion of the cardiac cycle to the end of the QRS complex across all leads. By varying $\RR$ and observing the resulting $\QT$ intervals, we evaluate the accuracy of the model in capturing this interdependency.

The results presented in \cref{fig:qt_as_rr_ecgs}, \cref{fig:qt_as_rr_curve} and \cref{tab:qt_vs_rr:r2} show a very strong agreement with the Fridericia model. This agreement suggests that the relationship between $\QT$ and $\RR$ intervals follows a defined but patient-dependent (and hence QRS-dependent) dynamic. Although clinical observations have previously shown this relationship, the fact that our model is able to regenerate it without being explicitly trained for this purpose is remarkable. This variation emphasizes the importance of conditioning on specific observations (such as the QRS complex) to reflect individual physiological differences. 
More generally, this observation suggests that our model reliably predicts ventricular repolarization (T-wave) for a given ventricular depolarization (QRS) and hence provides a way of computing a "normal" T-wave at the individual level. The possibilities associated with capability are numerous, especially for the detection of acute or chronic repolarization anomalies due to drugs, ionic disturbances or congenital and acquired diseases.

\begin{filecontents*}{data/goodness_of_fits_qt_vs_rr.csv}
,method,R2
0,Framingham,0.88 +/- 0.03
1,Bazett,0.47 +/- 0.04
2,Bazett with offset,0.98 +/- 0.00
3,Fridericia,0.94 +/- 0.02
4,Fridericia with offset,0.98 +/- 0.00
\end{filecontents*}
\DTLloaddb{qt_vs_rr_r2}{data/goodness_of_fits_qt_vs_rr.csv}
\begin{table}[]
\centering
\resizebox{.5\textwidth}{!}{
\begin{tabular}{|c|c|c|}%
    \hline
     Method & $R^2$ score & Expression \\
     \hline
     Framingham & 0.88 +/- 0.032 & $\QT=QTc + 0.154(1-\RR)$ \\
     Bazett & 0.47 +/- 0.042 & $\QT=QTc\sqrt{\RR}$\\
     Bazett with offset & 0.98 +/- 0.002 & $\QT=QTc_0\sqrt{\RR} + QTc_1$\\
     Fridericia & 0.94 +/- 0.022 & $\QT=QTc \sqrt[3]{\RR}$\\
     Fridericia with offset & 0.98 +/- 0.002 & $\QT=QTc_0 \sqrt[3]{\RR} + QTc_1$ \\\hline
    \end{tabular}}%
\caption{R2 score for each method for $\QT$ vs $\RR$.}
\label{tab:qt_vs_rr:r2}
\end{table}

\begin{minipage}[T]{0.4\textwidth}
    \begin{figure}[H]
    \centering
    \begin{tabular}{M{0.15\linewidth}@{\hspace{0\tabcolsep}}
    M{0.8\linewidth}@{\hspace{0\tabcolsep}}}
        \small QT(s) & \includegraphics[valign=C,width=\hsize]{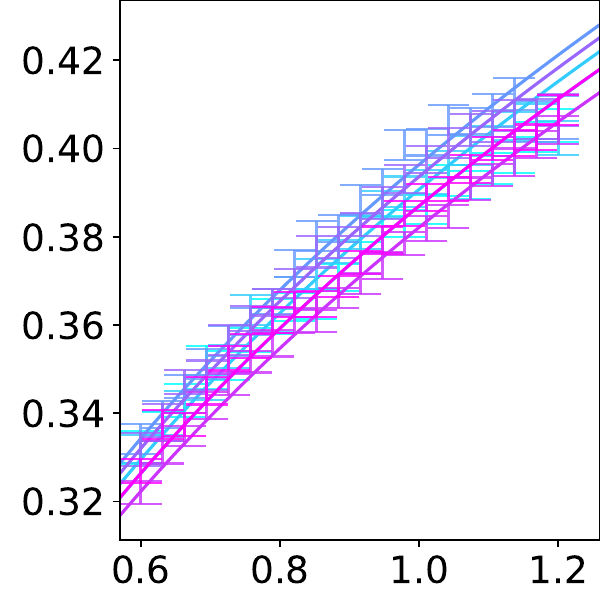} \\
        & \small RR(s) \\
    \end{tabular}
    \vspace{-.3mm}
    \caption{The figure displays the $\QT$ measured for each $\RR$ in the blue error bars, which consist of the $95\%$ CLT intervals obtained from the posterior samples. We show the Frederice fitted curves.}
    \label{fig:qt_as_rr_curve}
    \end{figure}
\end{minipage}
\hspace{0.04\textwidth}
\begin{minipage}[T]{0.56\textwidth}
    \begin{figure}[H]\includegraphics[valign=T,width=.95\hsize]{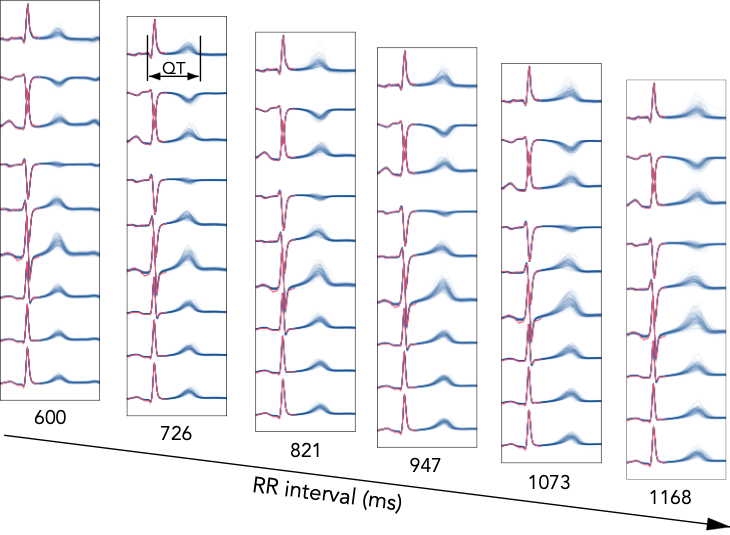}
        \centering
        \caption{The figure displays both the posterior ECG (in blue) and the conditioning ECG (in red) for different values of $\RR$.}
        \label{fig:qt_as_rr_ecgs}
    \end{figure} 
\end{minipage}

\subsubsection{Abnormality detection}
In this section, we illustrate the use of a DDGM specifically trained on healthy heart rhythms (as detailed in \Cref{sec:numerics:gen_model}) together with the conditional heartbeat generation algorithm (\algo). This combination serves as a novel tool for anomaly detection in cardiology. We focus on augmented limb leads (aVL, aVR, aVF), as described in \Cref{sec:DDGM:cond_gen}, which form our partial observation vector $\lmeas$. In contrast to the precordial leads close to the heart (V1-V6), these extended limb leads capture a broader electrical activity of the entire heart. They provide information about the overall orientation and geometry of the heart. However, due to their distance from the electrical sources, they tend to dilute the finer, localized electrical abnormalities that are usually more pronounced in the precordial leads.

Since our DDGM is capable of generating healthy heartbeats, we assume that the precordial leads should show normal patterns when reconstructed by \algo\ using the data from the extended limb leads, regardless of the underlying cardiac pathologies. This assumption is based on the fact that limb leads, which are less sensitive to local abnormalities, would not capture subtle pathologic changes.

Therefore, by superimposing the model-generated (posterior) ECG data on the actual ECG readings, we can effectively highlight discrepancies and thus detect cardiac abnormalities. This method promises to improve our ability to detect subtle but clinically significant ECG abnormalities by exploiting the contrast between the expected normality in the model output and the actual ECG readings.

In this section, we demonstrate our methodology using the diagnosis of myocardial infarction (MI) as an example. MI is a disease caused by the obstruction of coronary arteries, which leads to oxygen deprivation and the death of myocardial cells. This event causes the affected myocardial zones to lose their normal contractile function. In current clinical practice, MI is divided into two types: ST-elevation MI (STEMI) and non-ST-elevation MI (NSTEMI). The distinction between these two types is of crucial importance, even if it is often not recognizable on the electrocardiogram (ECG). It is critical to the choice of treatment, as certain therapies that are appropriate for one type may be contraindicated for the other~\cite{LAWTON2022e21}. Understanding these nuances in ECG interpretation is essential for accurate diagnosis and appropriate treatment planning.

\begin{figure}[]
\begin{minipage}[t]{0.43\textwidth}
\hspace{-1.5\tabcolsep}
\begin{tabular}{M{0.08\linewidth}@{\hspace{0\tabcolsep}} 
M{0.9\linewidth}@{\hspace{0\tabcolsep}}}
    \rotatebox[origin=c]{90}{Rescaled Distance\hspace{-4\tabcolsep}}   &
\includegraphics[valign=C,width=\hsize]{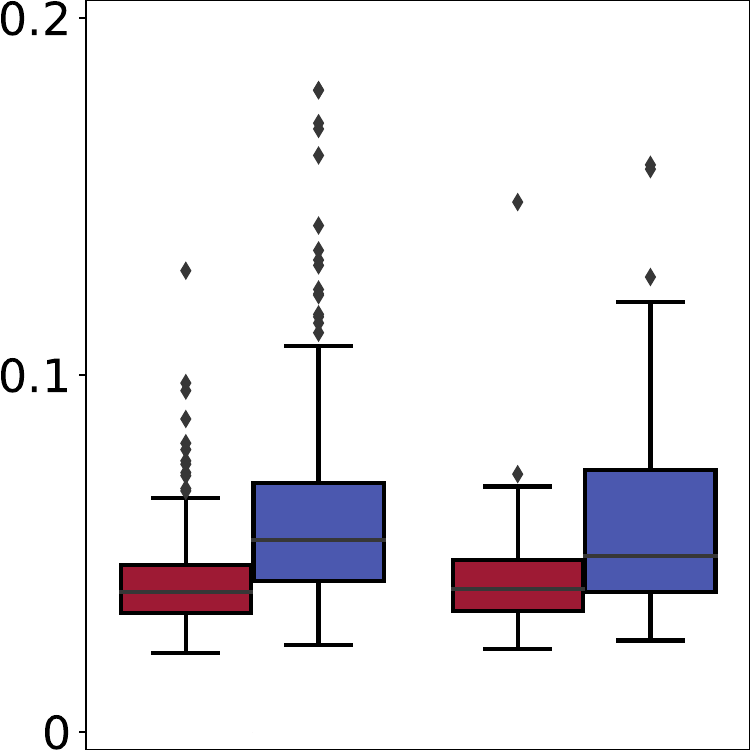}
  \\\addlinespace[-0.5ex]
  &
 \begin{tabular}{
M{0.5\linewidth}@{\hspace{0\tabcolsep}} 
M{0.37\linewidth}@{\hspace{0\tabcolsep}}}
 Male & Female
\end{tabular}
\end{tabular}
\caption{\small{Distribution of rescaled Mahalanobis distances between posterior distribution and target ECGs, averaged over tracks, for control (red) and MI (blue) sets.}}\label{fig:mahanalobis}
\end{minipage}%
    \quad
\begin{minipage}[t]{0.43\textwidth}
\hspace{-1.5\tabcolsep}
\begin{tabular}{M{0.08\linewidth}@{\hspace{0\tabcolsep}} 
M{0.9\linewidth}@{\hspace{0\tabcolsep}}}
    \rotatebox[origin=c]{90}{True Positive Rate\hspace{-4\tabcolsep}}   &
\includegraphics[valign=C,width=\hsize]{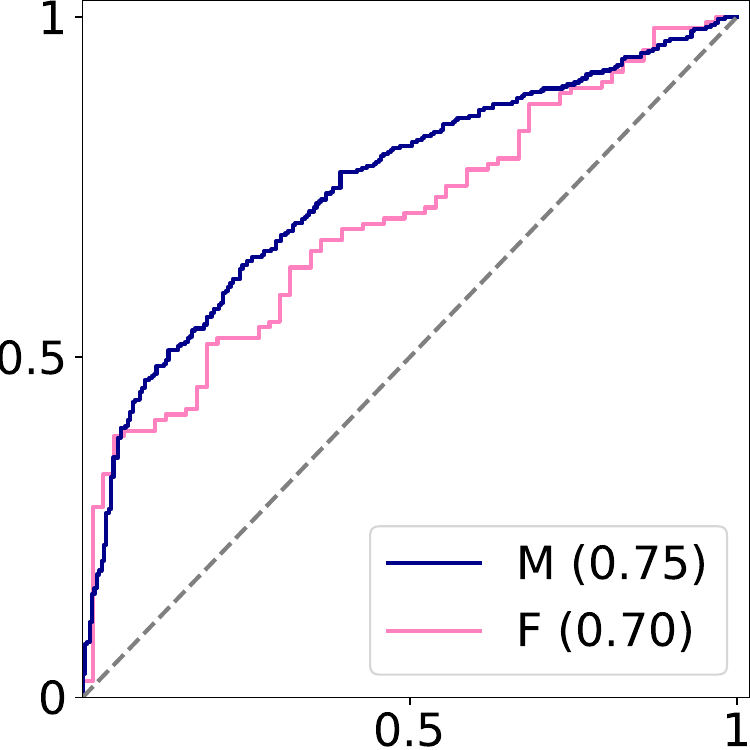}
  \\\addlinespace[-0.5ex]
  &
 False Positive Rate
\end{tabular}
\caption{\small ROC curve for classification between control and MI based on the distance between the posterior distribution and the real ECG.}\label{fig:ROC_MI}
\end{minipage}
\end{figure}

Considering that the incidence of MI mainly affects the elderly and that ECG characteristics vary according to age and gender even in healthy individuals, we carefully selected our control population from the test group. This selection ensures that the control group reflects the age and gender distribution of the MI patient cohort. Detailed information about the sampling procedure can be found in the appendix. 
For each patient in both the MI and healthy groups, we use \algo\ with $\npart=50$ particles to generate 100 samples from the distribution of $\state | \lmeas, \paramem_{\nmle}$. In this context, $\state$ represents a random variable modeling the single-beat ECG, $\lmeas$ refers to observations of limb-derived leads, and $\paramem_{\nmle}$ denotes the estimated variance of these observations by running \cref{alg:mle_noise_std} with $\nmle=10$ steps.
The rationale for the choice of $\npart=50$ is deferred to the appendix.
In our study, we assess the rescaled Mahalanobis distance, as described in \Cref{sec:numerics:applications:denoising}, to distinguish between control and MI patients. Consistent with our hypotheses, the Mahalanobis distance between posterior samples and actual ECG readings is generally smaller in control patients than in MI patients, as shown in \cref{fig:mahanalobis}. 
This finding is of critical importance as it suggests that MI patients can be accurately differentiated from control subjects based on this distance metric alone, which is also supported by the Receiver Operating Characteristic (ROC) analysis shown in \cref{fig:ROC_MI}.

\begin{figure}[]
\centering
\includegraphics[valign=T,width=.8\hsize]{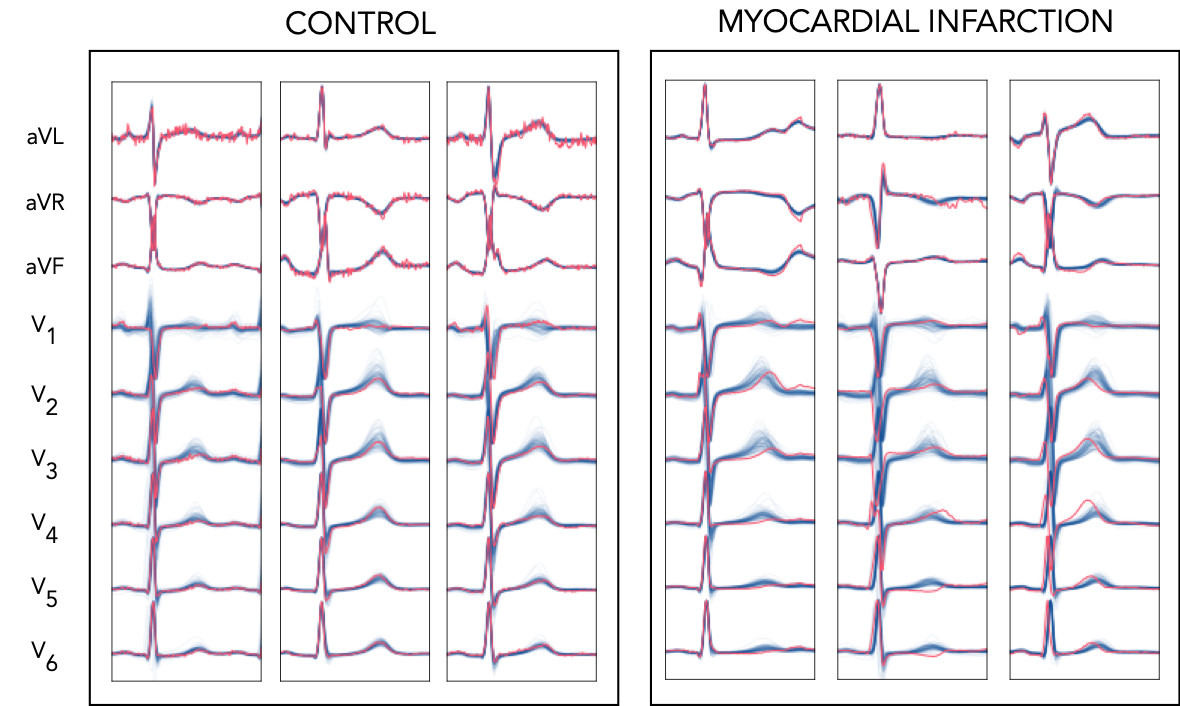}
\caption{The figure displays both the posterior ECG (in blue) and the patient's ECG (in red) for control and MI patients.}
\label{fig:ecg_examples_per_setting}
\end{figure}

Our method also enables the identification of ST-segment abnormalities. Specifically, ST-segment elevations or depressions are discerned when the ST segments of the algorithm-generated ECGs deviate from those of the actual ECGs - elevations are indicated when the generated segments are higher, and depressions when they are lower. This is exemplified in \cref{fig:ecg_examples_per_setting}, where the left MI example predominantly shows elevations, while the middle example displays depressions. Notably, in this middle case, ST segments in leads V2 and V3 are elevated, whereas those in V5 and V6 are depressed, showcasing the reciprocal nature of these changes.

One of the key strengths of our methodology lies in its ability to provide detailed and localized insights into the discrepancies between expected and actual ECG patterns. This level of precision, which focuses on specific areas of discrepancy, is beyond the capabilities of conventional black-box algorithms for detecting outliers. Our approach is characterized by its ability to identify the exact points of divergence in the ECG signal. This capability is not only a technical achievement, but also has significant clinical value. It improves ECG analysis by providing more detailed, actionable information that is particularly important for the diagnosis of  MI. This improved specificity allows for a deeper understanding of ECG changes associated with MI, which can lead to more accurate diagnoses and better patient outcomes

\section{Outlooks}
In this paper, we have focused on the technical aspects of our DDGM and provided an overview of possible initial applications. Beyond these initial applications, we expect this model to be useful for certain additional tasks:
\begin{itemize}
    \item Generation of 12-lead ECGs from single or 6-lead ECGs. This feature can be particularly valuable for transferring diagnostic algorithms trained on 12-lead ECGs to ECGs from smartwatches or connected devices.
    \item Identification of general repolarization abnormalities. The common measure of ventricular repolarization is the QT interval. Since ventricular repolarization depends on cellular parameters, but also on the activation sequence, we provide a way to investigate the cellular contribution to repolarization by "normalising" the T wave by the activation sequence. Possible applications include the diagnosis of long QT syndrome or other diseases that specifically alter repolarization.
    \item Generation of ECGs in certain situations. For this purpose, the model must be trained on more diverse data with labels according to pathologies. A concrete application example would be to train the model on patients with left bundle branch block so that we can detect ischemia in these patients for whom the ST segment elevation or depression criteria are not valid.
\end{itemize}
\section{Conclusion}
We have proposed a generative denoising diffusion model (DDGM) specifically trained on ECG signals from the healthy heart, focusing on accurately capturing the subtleties of ECG morphology. This model shows high fidelity in reproducing ECG data, which emphasizes its potential for clinical applications. Furthermore, we integrate this model with state-of-the-art techniques for solving linear inverse problems using DDGM priors to accomplish a number of critical clinical tasks, including denoising, missing lead recovery, QT interval prediction, and anomaly detection, without the need to retrain the original DDGM. In addition, these methods provide transparent, interpretable white-box algorithms. They allow a comparative visualization between the expected ECG result conditioned by certain clinical parameters and the patient's actual ECG. This approach not only increases diagnostic accuracy, but also contributes to a deeper understanding of the underlying cardiac conditions.

\section*{Acknowledgment}
The work of G.V.~Cardoso, R.~Dubois and J.~Duchateau is supported through the Investment of the Future grant, ANR-10-IAHU-04. The work of E.~Moulines and L.~Bedin is supported by the ANR-19-CHIA-0002, Inférence statistique, méthodes numériques et Intelligence Artificielle, SCAI. Part of this work has been carried out under the auspices of the Lagrange center for Mathematics and Computing, Paris. 

\bibliographystyle{plain}
\bibliography{bibliography}
\clearpage

\onecolumn
\section{Appendix}
\subsection{Additional theoretical results.}
In this section we prove two important aspects mentioned in \cref{sec:ddgm_ecg}. Namely, that the inference process matches the marginals of the forward process ($\revbridge{k|0}{\lstate_0}{\lstate_k} =  \fwmarg{t|0}{\lstate_k|\lstate_0}$) and that for a certain choice of weighting coefficients, \eqref{eq:opt_denoiser} consists of minimizing a certain $\kldivergence{}{}$, where for two densities $f, g$ we define
\begin{equation}
    \label{eq:kl_def}
    \kldivergence{f}{g} = \int \log\left(\frac{f(x)}{g(x)}\right) f(x)\rmd x \eqsp.
\end{equation}
This follows closely \cite{song2021denoising}, adapting it to our notation and to the variance exploding framework.
\begin{lemma}
    \label{lemma:inf_process_marginals}
    Let $\sequence{\infprocparam}[k][\nset]$ satisfy 
    $\infprocparam_{k}^2 \in [0,\stdexpl^2_k]$ for all $k \in \intvect{1}{\enddiff}$.
    Then
    \begin{equation*}
        \revbridge{k|0}{\lstate_0}{\lstate_k} =  \fwmarg{k|0}{\lstate_k|\lstate_0}\eqsp.
    \end{equation*}
\end{lemma}
\begin{proof}
    We proceed by induction. By definition, equality holds for $k = \enddiff$. Assume that for $k+1$ the equality holds. Then, note that
    \begin{equation*}
        \revbridge{k|0}{\lstate_0}{\lstate_k} = \int \revbridge{k|k+1, 0}{\lstate_{k+1}, \lstate_0}{\lstate_k} 
        \revbridge{k+1|0}{\lstate_0}{\lstate_{k+1}} \rmd \lstate_{k+1} = \int \gauss\left(\lstate_{k} ; \boldsymbol{\mu}_{k}(\lstate_0,\lstate_{k+1}), \infprocparam_{k}^2 \idm_{d} \right)\gauss\left(\lstate_{k+1}; \lstate_0, \stdexpl^2_{k+1}\idm \right) \rmd \lstate_{k+1} \eqsp,
    \end{equation*}
    with
    $\boldsymbol{\mu}_{k}(\lstate_0,\lstate_{k+1})= \lstate_0 + (\stdexpl^2_{k} / \stdexpl^2_{k+1} - \infprocparam_{k}^2 / \stdexpl^2_{k+1})^{1/2} (\lstate_{k+1} - \lstate_0) \eqsp.$
    By standard Gaussian conjugation formulas, we have that $\revbridge{k|0}{\lstate_0}{\lstate_k} = \gauss\left(\lstate_k; \lstate_0, \stdexpl^2_{k} \right)$, completing the proof.
\end{proof}
Note that taking $\infprocparam_k^2 = \frac{\stdexpl^2_{k}}{\stdexpl_{k+1}^2} \vestdker{k+1}^2$ yields $\revbridge{k|k+1, 0}{}{} = \realrevbridge{k|k+1, 0}{}{}$ where
    \begin{equation}
        \realrevbridge{k|k+1, 0}{\lstate_0, \lstate_{k+1}}{\lstate_k} \eqdef \frac{\fwtrans{k+1|k}{\lstate_{k}}{\lstate_{k+1}} \fwtrans{k|0}{\lstate_0}{\lstate_k}}{\fwtrans{k+1|0}{\lstate_0}{\lstate_{k+1}}}
        =\gauss(\lstate_k; \lstate_0 + \frac{\stdexpl^2_{k}}{\stdexpl^2_{k+1}}(\lstate_{k+1} - \lstate_0), \frac{\stdexpl^2_{k}}{\stdexpl^2_{k+1}} \vestdker{k+1}^2\idm) \eqsp.
    \end{equation}
    This shows that the inference process can be seen as a generalization of the forward noising process.
\begin{lemma}
    \label{lemma:kl_loss}
    Let $\mu(\lstate_{0:\enddiff}) = \datadistr(\lstate_0) \revbridge{1:\enddiff|0}{\lstate_0}{\lstate_{1:\enddiff}}$. Then, 
    \begin{equation}
        \kldivergence{}{}(\mu \parallel \bwoptmarg{\theta}{0:\enddiff}{}) = C
        + \sum_{k=1}^{\enddiff} \gamma_{k}^2 \expec{\state_0 \sim \datadistr, \epsilon \sim \normdist(0, \idm)}{\|\denoiser{0|k}[\theta](\state_{0} + \stdexpl_k \epsilon, \stdexpl_k) - \state_{0} \|^2}\eqsp,
    \end{equation}
    where $C$ is a constant independent of $\theta$ and 
    \begin{align*}
        \gamma_k^2 &= {\infprocparam_{k-1}^{-2}} {\left[1 - (\stdexpl^2_{k-1} / \stdexpl^2_{k} - \infprocparam_{k-1}^2 / \stdexpl^2_{k})^{1/2}\right]^2} \quad \text{for} \quad k > 1 \eqsp,\\
        \gamma_1^2 &= \infprocparam^{-2}_0 \eqsp.
    \end{align*}
\end{lemma}

\begin{proof}
    In this proof, we treat every constant not depending on $\theta$ as $C$. Note that the actual value of $C$ can change from a line to the other. We start by rewriting 
    \begin{align*}
        \kldivergence{}{}(\mu \parallel \bwoptmarg{0:\enddiff}{\theta}{}) &= \int \log \left(\frac{\datadistr(\lstate_0) \revbridge{1:\enddiff|0}{\lstate_0}{\lstate_{1:\enddiff}}}{\bwmarg{0:\enddiff}{\lstate_{0:\enddiff}}}\right) \datadistr(\lstate_0) \revbridge{1:\enddiff|0}{\lstate_0}{\lstate_{1:\enddiff}} \rmd \lstate_{0:\enddiff}\\ 
        &=\sum_{k=1}^{\enddiff-1} \int \log \left(\frac{\revbridge{k|k+1, 0}{\lstate_0, \lstate_{k+1}}{\lstate_k}}{\bwtrans{\theta}{k|k+1}{\lstate_{k+1}}{\lstate_k}}\right) \revbridge{k|k+1, 0}{\lstate_0, \lstate_{k+1}}{\lstate_k} \revbridge{k+1|0}{\lstate_0}{\lstate_{k+1}} \datadistr(\lstate_0) \rmd \lstate_{0, k, k+1}\\
        &+ \int \log \left(\frac{\datadistr(\lstate_0) \revbridge{1|0}{\lstate_0}{\lstate_1}}{\bwtrans{\theta}{0|1}{\lstate_1}{\lstate_0}}\right) \datadistr(\lstate_0) \revbridge{1|0}{\lstate_0}{\lstate_1} \rmd \lstate_{0:1} + C \\
        &= \sum_{k=1}^{\enddiff-1} \int \kldivergence{}{}(\revbridge{k|k+1, 0}{\lstate_0, \lstate_{k+1}}{\cdot} \parallel {\bwtrans{\theta}{k|k+1}{\lstate_{k+1}}{\cdot}}) \fwmarg{k+1|0}{\lstate_0}{\lstate_{k+1}} \datadistr(\lstate_0) \rmd \lstate_{0, k+1}\\
        &- \int \log \bwtrans{\theta}{0|1}{\lstate_1}{\lstate_0} \datadistr(\lstate_0) \revbridge{1|0}{\lstate_0}{\lstate_1} \rmd \lstate_{0:1} + C \eqsp,
    \end{align*}
    where $C$ is a constant that does not depend on $\theta$.
    We know that 
    \begin{equation*}
        \kldivergence{\gauss(\mu_1, \sigma_1^2 \idm)}{\gauss(\mu_2, \sigma_2^2 \idm)} = 2^{-1} \left[2d \log(\sigma_2 / \sigma_1) - d  + d (\sigma_1 / \sigma_2)^2 + \|\mu_2 - \mu_1\|^2 / \sigma_2^2\right] \eqsp,
    \end{equation*}
    thus
    \begin{align*}
    \kldivergence{}{}(\revbridge{k|k+1, 0}{\lstate_0, \lstate_{k+1}}{\cdot} \parallel {\bwtrans{\theta}{k|k+1}{\lstate_{k+1}}{\cdot}})& = {\infprocparam_k^{-2}} {\left[1 - (\stdexpl^2_{k} / \stdexpl^2_{k+1} - \infprocparam_{k}^2 / \stdexpl^2_{k+1})^{1/2}\right]^2} \|\denoiser{0|k+1}[\theta](\lstate_{k+1}) - \lstate_0 \|^2 \eqsp.
    \end{align*}
    Note also that
    \begin{equation*}
         \log \bwtrans{\theta}{0|1}{\lstate_1}{\lstate_0} = -\infprocparam_0^{-2} \|\denoiser{0|1}[\theta](\lstate_{1}) - \lstate_0\|^2 + C \eqsp.
    \end{equation*}
    The proof is finished by \cref{lemma:inf_process_marginals}.
\end{proof}

\twocolumn
\subsection{Dataset}
\begin{table}[H]
    \centering
    \resizebox{.48\textwidth}{!}{
    \begin{tabular}{|c|c|c|c|c|}%
        \hline
          & Train & CV & Test & MI \\
         \hline
         All (patients) & 22580 & 2723 & 2864 & 468 \\
        Male (patients) & 11722 & 1399 & 1497 & 343 \\
        Female (patients) & 10858 & 1324 & 1367 & 125 \\
        All (beats) & 214460 & 25694 & 27221 & 44911\\
         Mean (beats) & 9.5 +/- 0.1 &  9.4 +/- 0.2& 9.5 +/- 0.2& 96 +/- 5\\
         \hline
        \end{tabular}}%
    \caption{Distribution of patients, gender and number of recorded beats among train, test and MI sets.}
    \label{tab:dataset_splits}
\end{table}

\subsection{Number of particles}
As the number of particles, denoted as $N$, increases, we observe a corresponding decrease in the discrepancy between the target posterior distribution and the distribution of particles generated by \algo. A critical question arises: what is the optimal value for $N$ that strikes a balance between accuracy and computational efficiency? To approach this question, we first selected a patient from the test dataset and used \algo\ to generate $10^3$ samples with a high particle count of $N=10^4$. We consider these samples as our reference representing the target posterior distribution.

We then generated $10^3$ samples with \algo\ for different values of $N$ and calculated the Earth Mover's Distance (EMD) relative to the reference samples. This process helps us to evaluate the convergence of the distribution generated by the algorithm to the posterior as $N$ varies. \Cref{fig:emd_diffusion} illustrates the relationship between $N$ and the EMD. From this analysis, $N=50$ provides an effective equilibrium that provides a reasonable approximation to the posterior distribution while ensuring manageable inference times.
\begin{figure}[H]
\centering
\begin{tabular}{M{0.1\linewidth}@{\hspace{0\tabcolsep}} M{0.4\linewidth}@{\hspace{0\tabcolsep}}}
  \STAB{\rotatebox[origin=c]{90}{\small EMD}}
  &  \includegraphics[width=\hsize]{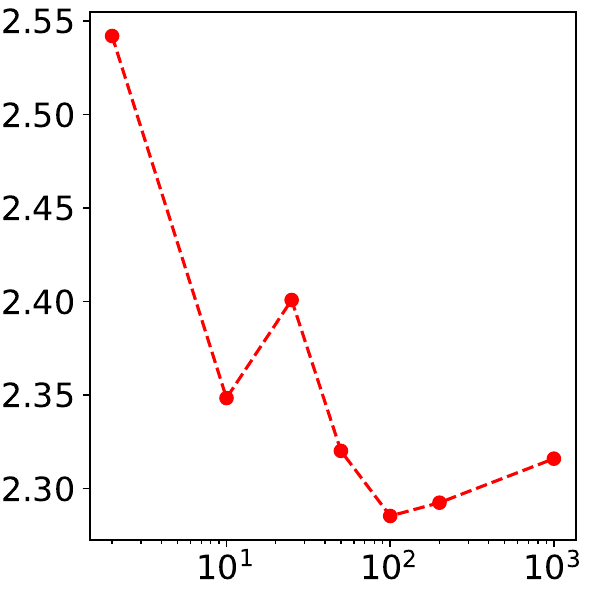}\\\addlinespace[-1ex]
  & \small $N$
\end{tabular}
\vspace{-.3mm}
\caption{EMD distance between $1000$ samples from \algo\ with $N$ particles and $1000$ samples of \algo\ with $10^5$ particles, that is considered the standard samples.}
\label{fig:emd_diffusion_per_patient}
\end{figure}
\subsection{Balancing age-by-sex distribution}
To assess the ability to detect abnormalities in MI ECGs compared with control healthy ECGs, we need to sample healthy ECGs with the same age distribution as MI patients. We approximate the age-by-sex density in the MI population with a Gaussian kernel and a bandwidth of 2 years. Next, we sample a subset of healthy ECGs based on this distribution. Since the patients must be distinct, the number of sampled ECGs is much smaller than the total number of healthy patients in the test set, but nevertheless, the number of sampled ECGs remains of the same order of magnitude as the size of the MI set. Indeed, we managed to obtain 63 female control ECGs and 274 male control ECGs, while the MI set contains 125 female ECGs and 343 male ECGs.

\subsection{Generated ECGs}
\begin{figure}[H]
\centering
\begin{tabular}{M{0.11\linewidth}@{\hspace{-2\tabcolsep}} 
M{0.87\linewidth}@{\hspace{0\tabcolsep}}}
\addlinespace[-2mm]
    \adjustbox{valign=C}{
    \begin{tabular}{c}
    \\\addlinespace[1mm]
         aVL\\\addlinespace[0.85mm]
         aVR\\\addlinespace[0.85mm]
         aVF\\\addlinespace[0.85mm]
         V1\\\addlinespace[0.85mm]
         V2\\\addlinespace[0.85mm]
         V3\\\addlinespace[0.85mm]
         V4\\\addlinespace[0.85mm]
         V5\\\addlinespace[0.85mm]
         V6
    \end{tabular}}
    &
\begin{tabular}{
M{0.15\linewidth}@{\hspace{0\tabcolsep}} 
M{0.15\linewidth}@{\hspace{0\tabcolsep}}
M{0.15\linewidth}@{\hspace{0\tabcolsep}} 
M{0.15\linewidth}@{\hspace{0\tabcolsep}}
M{0.15\linewidth}@{\hspace{0\tabcolsep}} 
M{0.15\linewidth}@{\hspace{0\tabcolsep}}}
    \includegraphics[valign=T,width=\hsize]{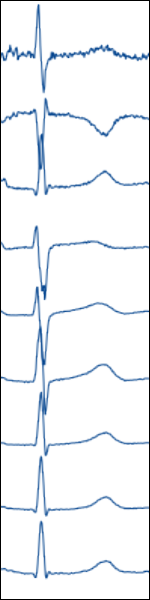} &
  \includegraphics[valign=T,width=\hsize]{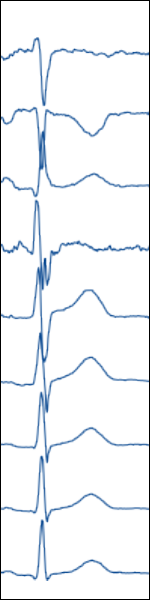} &
  \includegraphics[valign=T,width=\hsize]{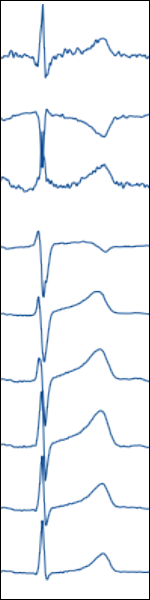} &
  \includegraphics[valign=T,width=\hsize]{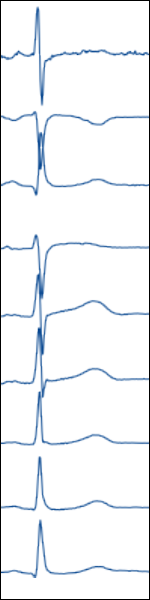} &
  \includegraphics[valign=T,width=\hsize]{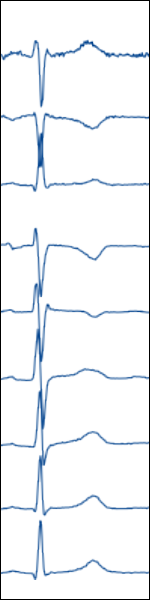} &
  \includegraphics[valign=T,width=\hsize]{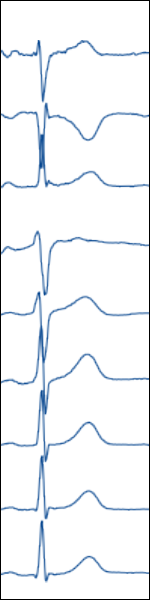}
\\\addlinespace[-1ex]
\end{tabular}
\end{tabular}
\vspace{-.3mm}
\caption{ECGs generated from the DDGM with random conditioning features sampled from the test dataset.}
\label{fig:ecg_examples}
\end{figure}

We display in \cref{fig:ecg_examples} some generated ECGs from the DDGM. The conditioning features are taken randomly from the test dataset.

\subsection{Architecture details}
We implement a very close architecture to \cite{Karras2022edm} and available at \url{https://github.com/NVlabs/edm} as well as training procedure. The main difference is that we replaced the 2D convolutional layers by 1D ones in every UNet. The final network use the following parameters:
\begin{itemize}
    \item First embedding dimension: $192$,
    \item Number of Unet blocks per resolution: 2,
    \item Number of resolutions: 1,
    \item Dropout probability $0.10$.
\end{itemize}
For the training, the following configuration was used:
\begin{itemize}
    \item learning rate: $10^{-4}$,
    \item Number of epochs: $10^4$,
    \item Batch Size: $1024$,
    \item Exponential moving average coefficient: $0.9999$.
\end{itemize}
For the (forward diffusion) we used the following parameters:
\begin{itemize}
    \item $\sigma_{\operatorname{min}} = 2 \times 10^{-4}$,
    \item $\sigma_{\operatorname{max}} = 80$,
    \item $\sigma_{\operatorname{data}} = 0.5$,
    \item Importance law of $\sigma$ for training: $\operatorname{Log}\normdist(-1.2, 1.2^2\idm)$.
\end{itemize}

\newpage

\end{document}